\documentclass[superscriptaddress,nobibnotes,amsmath,amssymb,notitlepage,twocolumn,pra,longbibliography]{revtex4-2}

\usepackage{bm,braket}
\usepackage[toc,page]{appendix}
\usepackage{comment}
\usepackage{dcolumn}

\usepackage{amsfonts}

\usepackage{graphicx,color,hyperref}
\usepackage[caption=false]{subfig}
\hypersetup{colorlinks=true, linkcolor=blue, citecolor=blue, urlcolor=blue} 

\graphicspath{{./img/}}

\newcommand{\beq}[1]{\begin{equation}\label{#1}}
\newcommand{\eep}{\;.\end{equation}}
\newcommand{\eec}{\;,\end{equation}}
\newcommand{\eeq}{\end{equation}}

\newcommand*\dd{\mathop{}\!\mathrm{d}} %differential d

\newcommand{\om}{\omega}

\newcommand{\Om}{\Omega}

\DeclareMathAlphabet{\mathcal}{OMS}{cmsy}{m}{n} % Changes font for mathcal but leaves the rest of the math fonts in Times.

\renewcommand{\vec}[1]{{\bf #1}}

\newcommand{\kv}{\vec{k}}

\usepackage{amsmath}
\usepackage{amssymb}
\usepackage{xcolor}
\usepackage{bbm}
\usepackage{physics}
\usepackage{float}
\usepackage{graphicx}
\usepackage{dcolumn} % Align table columns on decimal point
\usepackage{bm} % bold math
\usepackage{siunitx}
\usepackage{enumitem}  % to use (i) etc for enumerate lists
\makeatletter
\renewcommand*{\fnum@figure}{{\normalfont\bfseries \figurename~\thefigure}}
\makeatother

\allowdisplaybreaks

\definecolor{orange}{rgb}{1,0.5,0}

\newcommand{\sect}[1]{\vspace{0.3em}{\it #1.}---}

\graphicspath{{./img/}}

\DeclareMathAlphabet{\mathcal}{OMS}{cmsy}{m}{n} % Changes font for mathcal but leaves the rest of the math fonts in Times.

\newcommand{\intBZ}{\int_{\text{BZ}}} % Integration over BZ

\makeatletter
\newcommand{\specificthanks}[1]{\@fnsymbol{#1}}% Inserts a specific \thanks symbol
\makeatother

\begin{document}

\preprint{APS/123-QED}

\title{Quantum geometric bounds in spinful systems with trivial band topology}

\author{Wojciech J. Jankowski}
\email{wjj25@cam.ac.uk}
%\thanks{}
\affiliation{TCM Group, Cavendish Laboratory, Department of Physics, J J Thomson Avenue, University of Cambridge, Cambridge CB3 0HE, United Kingdom}

\author{Robert-Jan Slager}
%\email{rjs269@cam.ac.uk}
\affiliation{Department of Physics and Astronomy, University of Manchester, Oxford Road, Manchester M13 9PL, United Kingdom}
\affiliation{TCM Group, Cavendish Laboratory, Department of Physics, J J Thomson Avenue, University of Cambridge, Cambridge CB3 0HE, United Kingdom}

\author{Gunnar F. Lange}
\email{g.f.lange@fys.uio.no}
%\thanks{}
\affiliation{Department of Physics, University of Oslo, N-0316 Oslo, Norway}
\affiliation{Centre for Materials Science and Nanotechnology, University of Oslo, N-0316 Oslo, Norway}

\date{\today}

\begin{abstract}
    We derive quantum geometric bounds in spinful systems with spin topology characterized by a single $\mathbb{Z}$ index protected by a spin gap. Our bounds provide geometric conditions on the spin topology, distinct from the known quantum geometric bounds associated with Wilson loops and nontrivial band topologies. As a result, we obtain broader bounds in time-reversal symmetric systems with a nontrivial $\mathbb{Z}_2$ index and also  bounds in systems with a trivial $\mathbb{Z}_2$ index, where the quantum metric should be otherwise unbounded. We benchmark these findings with first-principles calculations in elemental bismuth realizing a nontrivial even spin-Chern number. Moreover, we connect these bounds to optical responses and show their robustness in the presence of disorder within a real space marker formulation, demonstrating that spin-resolved quantum geometry is observable in realistic experimental settings of impure materials. Finally, we connect spin bounds to quantum Cram\'er-Rao bounds that are central to quantum metrology, showing that elemental Bi and other spin-topological phases hold promises for topological free fermion quantum sensors.
\end{abstract} 

%\keywords{Suggested keywords}

\maketitle

\sect{Introduction}
Topological materials~\cite{Rmp1,Rmp2,Rmp3, Slager_2013,Kruthoff2017, Clas4, rjs_translational,Clas5, Fu_2011} constitute intriguing new phases of matter, as they display robust windings in the phases
of Bloch eigenvectors, which have direct physical consequences~\cite{Niu1985, Haldane1988}. More recently, it has been shown that these windings, formally characterized by the Wilson loops, also have implications for the 
Riemannian geometry of the eigenvectors~\cite{provost1980riemannian, Marzari1997, resta_2011_metric, Ahn2020, Ahn2021, Torma2023, Bouhon2023geometric}, placing robust lower bounds on the metric of the eigenvectors, integrated
across the Brillouin zone (BZ). Intuitively, this is because the winding of the states cannot be trivialized without closing the energy gap; i.e.
there must be some minimal momentum-space dependence in the eigenstates due to the nontrivial topology~\cite{Rahul2014, Xie2020, Palumbo2021, Bouhon2023geometric, jankowski2023optical, Kwon2024, jankowskiPRB2024Hopf, yu2024Z2}. This has important implications for optics, as the optical responses of materials depend on the quantum metric~\cite{Souza2000, Ahn2021}. Moreover, the geometry of states is of key importance for quantum metrology~\cite{Yu2022, SciPostMera2022, Yu2024, Wahl2025}
and quantum sensing~\cite{Degen2017, Sarkar2022}, as the metric is related to the quantum Fisher information~\cite{Liu2020}, which places a bound on the
sensitivity of sensors via the Cram\'er-Rao bound~\cite{Liu2020, Yu2022, SciPostMera2022,Yu2024, Wahl2025}. 

\begin{figure}
    \centering
    \def\svgwidth{0.95\linewidth}
    \includegraphics[width=\linewidth]{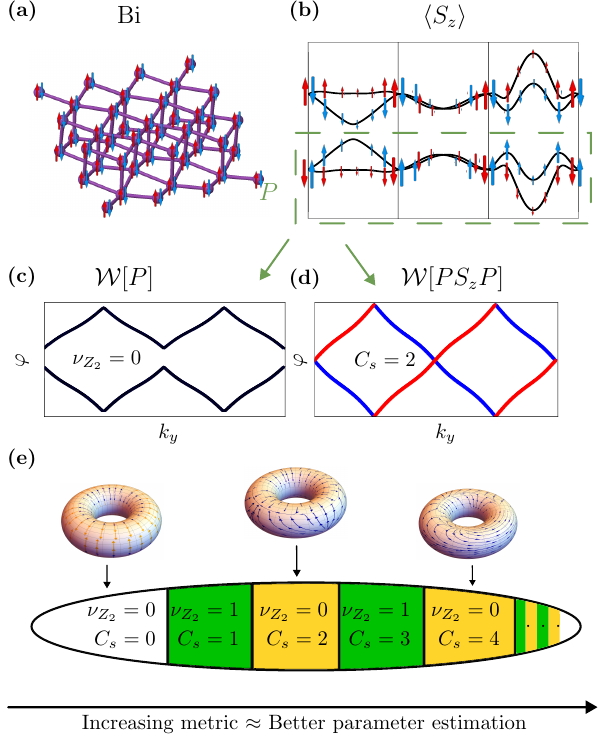}
    \caption{Illustration of the main results of this work. \textbf{(a)} For a non-magnetic material such as ultrathin bismuth, the energy bands may not have a definite spin value, due to spin-orbit entanglement, as illustrated schematically in panel~\textbf{(b)}, but the spin structure of energetically isolated bands (green dashed box) with projector $P$ can still be defined via the projected spin operator ($PS_zP$). The electronic topology, captured by the Kane-Mele invariant $\nu_{\mathbb{Z}_2}$ is only protected for odd relative Wilson loop windings of the bands in $P$, $\mathcal{W}[P]$, while even relative Wilson loop windings of the bands can gap out \textbf{(c)}. As long as the energy and spin gap (defined in Sec.~I~of the Supplemental Material~\cite{SI}) stay open, however, the spin topology, captured by the spin-Chern number $C_s$, is well-defined, and the spin-Wilson loop winding shown in panel~\textbf{(d)} cannot gap. This has direct consequences for the lower bound of the integrated quantum metric, which is crucial for quantum metrology. Previously, only phases with $\nu_{\mathbb{Z}_2} = 1$ [green in panel~\textbf{(e)}] had a known lower bound. We formulate a new bound for phases with even $C_s$ [orange in panel~\textbf{(e)}], and extend the bound for phases with odd $C_s>1$.}
    \label{fig:overview_fig}
\end{figure}

The paradigmatic example of a topological phase in time-reversal symmetric materials is the quantum spin Hall phase~\cite{Kane2005a, Kane2005b}.
This phase is realized in real materials, perhaps most notably in (111) terminations of elemental bismuth (Bi)~\cite{MurakamiQSH,111_Bi_exp_1,111_Bi_exp_2}. In general the topological invariant
associated with this phase is $\mathbb{Z}_2$-valued, and given by the Kane-Mele index $\nu_{\mathbb{Z}_2} \in \{0,1\}$ \cite{Kane2005b, FuKaneIndex}. It was realized early on,
however, that a $\mathbb{Z}$-valued invariant $C_s$, the spin-Chern number, could also be defined for such phases, giving
rise to the concept of spin topology. As long as the spin-orbit entanglement is moderately large, i.e., the spin gap is preserved~\cite{Prodan2009}, this remains
well-defined, but for large spin-orbit entanglement, i.e., when the spin gap is closed, only $\nu_{\mathbb{Z}_2} = C_s\ \mathrm{mod}\ 2$ is robust. This
has led to the realization that the topological $\nu_{\mathbb{Z}_2} = 1$ phase in elemental Bi can
actually be characterized by a higher spin-Chern number $C_s = 3$~\cite{Peng2024}. Most intriguingly, there also
exists a phase of elemental Bi with $C_s = 2$, with a~trivial $\nu_{\mathbb{Z}_2} = 0$ invariant (see Fig.~\ref{fig:overview_fig}).
This phase shows interesting behavior, including weakly gapped edge states, but as there is generically
no strict quantization of a bulk Wilson-loop invariant and no exact bulk-boundary correspondence, it is difficult to pin down a clear experimental signature of this phase. Here, we ask the question what implications this spin topology has for the quantum metric. We provide a lower bound on the metric by the spin-topological invariant, $C_s \in \mathbb{Z}$, the tightness of which depends on the degree of spin-orbit entanglement. For the
highly entangled case, this bound generalizes the recently discovered bound for the $\nu_{\mathbb{Z}_2}$ invariant~\cite{yu2024Z2}, and can be larger than it. Moreover, for the $\nu_{\mathbb{Z}_2} = 0$ case no metric bounds were previously identified. Thus, in this Letter, we describe the first lower bound for such phases. Our bound is a geometric bound purely due to the spin topology and associated winding of the projected spin operator ($PS_zP$) eigenvalues~\cite{Lange2023, Hernandez2024, Lin2024}, which holds even in the presence of a trivial band topology characterized by a trivial Wilson loop winding. We further show that our result culminates in  $\mathbb{Z}$ bounds on optical responses and band gaps in spin-topological materials. Having first detailed the spin topology (valid also in systems with trivial Wilson loops) in the context of which we introduce a spin-resolved quantum geometry, we retrieve  spin-topological bounds in ultrathin Bi [see~Fig.~\ref{fig:overview_fig}(a)] with $\nu_{\mathbb{Z}_2}=0$ and in few-band effective Hamiltonians that can be engineered and used for spin-topological quantum sensing and optics.

\sect{Spin topology}
We consider 2D spinful time-reversal symmetric ($\mathcal{T}^2 = -1$), periodic systems described by Bloch Hamiltonians $H(\boldsymbol{k})$. In a tight-binding basis with spin orbitals $|\alpha\rangle\otimes|\sigma\rangle$, such Hamiltonians can be written as
\begin{equation}
    H(\boldsymbol{k}) = \sum_{\alpha,\beta=1}^{N_{\mathrm{orb}}}\sum_{\sigma,\sigma' = \uparrow\downarrow} H_{\alpha\sigma,\beta\sigma'}(\boldsymbol{k})[|\alpha\rangle\otimes|\sigma\rangle][\langle \beta|\otimes \langle \sigma'|].
\end{equation}
The spin operator along $\hat{n}$ associated with this basis is then given by $S_{\hat{n}}=1_{\mathrm{orb}}\otimes \hat{n}\cdot \vec{\sigma}$. If $[H(\boldsymbol{k}),S_{\hat{n}}]=0$, then the eigenstates of $H(\boldsymbol{k})$ are spin eigenstates, and can be labeled as $|\tilde{u}_{n\boldsymbol{k}}^{\sigma}\rangle$, where $n\in\{1,\dots,N_{\mathrm{orb}}\}$. When this can be done for all $\boldsymbol{k}$ (i.e., there is no entanglement between spin and orbit degrees of freedom), one can define the multiband spin-Berry curvature:
\begin{equation}
\Omega_{ij,nm}^{\sigma}(\boldsymbol{k}) =i[\langle \partial_{k_i}\tilde{u}_{n\boldsymbol{k}}^{\sigma}|\partial_{k_j}\tilde{u}_{m\boldsymbol{k}}^{\sigma}\rangle - \mathrm{c.c}].
\end{equation}
Tracing over the occupied bands with spin $\sigma$, ${N_{\mathrm{occ}}^{\sigma} = N_{\mathrm{occ}}/2}$, this will be quantized when integrated over a 2D Brillouin zone (BZ), giving rise to the spin-resolved Chern numbers ${C^{\sigma}_s= \frac{1}{2\pi} \intBZ \dd \textbf{k}~ \text{Tr}~\Om^{\sigma}_{xy} \in \mathbb{Z}}$, with the overall spin-Chern number $C_s = \frac{1}{2}(C_s^{\uparrow}-C_s^{\downarrow})$. Time-reversal symmetry ensures that the Chern numbers of the opposite spin bands satisfy: ${C_s^{\uparrow} = -C_s^{\downarrow}}$~\cite{Sheng2006,Prodan2009}. When spin is not conserved, one can instead formulate the equivalent quantities in terms of the eigenstates $\ket{u_{n\kv}^{\sigma}}$ of the \textit{projected} spin operator $S_P^z = PS_zP$ [see~Fig.~\ref{fig:overview_fig}(b)], where $P$ is the projector onto the occupied bands. These eigenstates will not generically be energy eigenstates when spin is not a good quantum number, but $P = P_{\uparrow}+P_{\downarrow}$, where $P_{\sigma}=\sum_n^{N_{\text{occ}}^{\sigma}}\ket{u_{n\kv}^{\sigma}}\bra{u_{n\kv}^{\sigma}}$ projects onto the top and bottom $N^{\sigma}_{\text{occ}}$ states respectively; for a review of projected spin operators see Ref.~\cite{Lin2024} and Sec.~I~of the Supplemental Material~\cite{SI}. As we show in the following, formulating the spin-Berry curvature $\Om^{\sigma}_{xy}$ in terms of projected spin eigenstates $\ket{u_{n\kv}^{\sigma}}$ provides a local bound on the quantum metric, which holds at every point of momentum space, even when spin is not a good quantum number.

%Final figure: Scaling in bismuth 
%Metric, Curvature, Energy Gap, Spin Gap

\sect{Spin-resolved quantum geometry}
We now define the spin-resolved quantum geometry, which is fully characterized by a spin-resolved quantum geometric tensor (QGT), termed spin-QGT:
\beq{}
    Q^\sigma_{ij} \equiv \sum^{N^{\sigma}_\text{occ}}_{n} \bra{\partial_{k_i} u^\sigma_{n\kv}} 1-P_\sigma \ket{\partial_{k_j} u^\sigma_{n\kv}} = g^\sigma_{ij} - \frac{i}{2} \Om^\sigma_{ij},
\eeq
with $\sigma = \uparrow, \downarrow$, and spin-resolved quantum metric (spin~metric) $g^\sigma_{ij}$. The spin-QGT definitionally satisfies a positive semidefiniteness condition $\text{det}~Q^{\sigma}_{ij} > 0$, and as a consequence, we derive the spin-geometric bound (see Sec.~II of the Supplemental Material~\cite{SI}):
\beq{eq:localbound}
    \frac{1}{2} \text{Tr}~g^{\sigma}_{ij} \geq \sqrt{\text{det}~g^{\sigma}_{ij}} \geq \frac{|\text{Tr}~\Om^\sigma_{xy}|}{2}.
\eeq
%.
When spin-$z$ is a good quantum number, we further show that the quantum metric $g_{ij} = \frac{1}{2}\text{Tr}~[\partial_{k_i}P\partial_{k_j}P]= g^{\uparrow}_{ij} + g^{\downarrow}_{ij}$~\cite{provost1980riemannian} (see Sec.~III of the Supplemental Material~\cite{SI}). On combining with the spin-resolved bounds and integrating, one obtains~\cite{Fu2024}: 
\beq{eq:invbound}
    \frac{1}{2\pi} \intBZ~\dd^2\kv~(g_{xx} + g_{yy}) \geq |C^\uparrow_s| + |C^\downarrow_s| \equiv S_{\lambda=0} \in \mathbb{Z}.
\eeq
We next focus on the case where spin-$z$ is no longer a good quantum number, i.e., $[H,S_z] \neq 0$, but the spin gap is preserved~\cite{Lin2024, Bouhon2019}. In this case, utilizing the Cauchy-Schwarz inequality (see Sec.~III of the Supplemental Material~\cite{SI}) allows us to derive a fundamental bound with a suppression by the number of occupied bands, $N_\text{occ}$, due to the spin-orbit entanglement (see Sec.~IV of the Supplemental Material~\cite{SI}),
\beq{}\label{eq:bound_with_SOC}
    \frac{1}{2\pi} \intBZ~\dd^2\kv~(g_{xx} + g_{yy}) \geq \frac{|C^\uparrow_s| + |C^\downarrow_s|}{N_\text{occ}} \equiv S_{\lambda\neq0} \in \mathbb{Q},
\eeq
where on the right, we have a rational number $S_{\lambda \neq 0}$. Intuitively, the suppression by the factor $N_\text{occ}$ arises due to the mixing of all occupied bands with spin-orbit coupling (SOC), while preserving the winding of the spin-Wilson loop realized by the $PS_zP$ eigenstates. For odd spin-Chern numbers, our bound is consistent with the $\mathbb{Z}_2$~bound~of Ref.~\cite{yu2024Z2}, yet realizes a possible generalization to tighter inequalities for higher odd spin-Chern numbers, e.g., $|C_s| = 3$, depending on the ratio of $|C_s|$ and $N_{\mathrm{occ}}$. Furthermore, our bound provides for a more general fundamental quantum geometric condition, even if the $\mathbb{Z}_2$-invariant is trivial, i.e., when $C^\sigma_s \equiv 0~\text{mod}~2$. Namely, for even spin-Chern numbers Eq.\,\eqref{eq:bound_with_SOC} is the first topological bound due to spin topology, yet accompanied by trivial band topology. We further detail direct consequences of the spin-geometric bounds, which we further demonstrate to apply to spin-topological materials such as ultrathin Bi.

\sect{Spin-resolved geometry in optics} We proceed to show that as a direct physical consequence, the spin-resolved quantum geometry has direct manifestations in optical responses of materials with nontrivial spin-resolved topology, such as bismuth central to this work. The integral of the quantum metric amounts to the optical weight due to the optical conductivities $\sigma_{xx}(\om)$, $\sigma_{yy}(\om)$ at frequencies $\om$~\cite{Souza2000, Fu2024}, which therefore satisfies the spin-topological bounds,
\beq{}
    \int^\infty_0~\dd\om~\frac{\text{Re} [\sigma_{xx}(\om) + \sigma_{yy}(\om)]}{\om} \geq \frac{e^2}{2\hbar}  S_\lambda.
\eeq
We further retrieve an optical bound on the gap of spin-topological systems (see Sec.~V of the Supplemental Material~\cite{SI}),
\beq{}
    E_g \leq \frac{2\pi  n \hbar^2}{m S_\lambda}, 
\eeq
where $n$ is electron charge density, $m$ is electron mass, and $E_g$ is the energy band gap. Our bounds hold also in the $\mathbb{Z}_2$-trivial systems, i.e., where the Kane-Mele invariant is vanishing $\nu_{\mathbb{Z}_2} = 0$, placing our bound beyond the findings of Refs.~\cite{Fu2024, yu2024Z2}. We, moreover, validate our quantum geometric bounds in real material settings of ultrathin Bi, which we retrieve below.

\begin{figure}
    \centering
    \def\svgwidth{0.95\linewidth}
    \includegraphics[width=\linewidth]{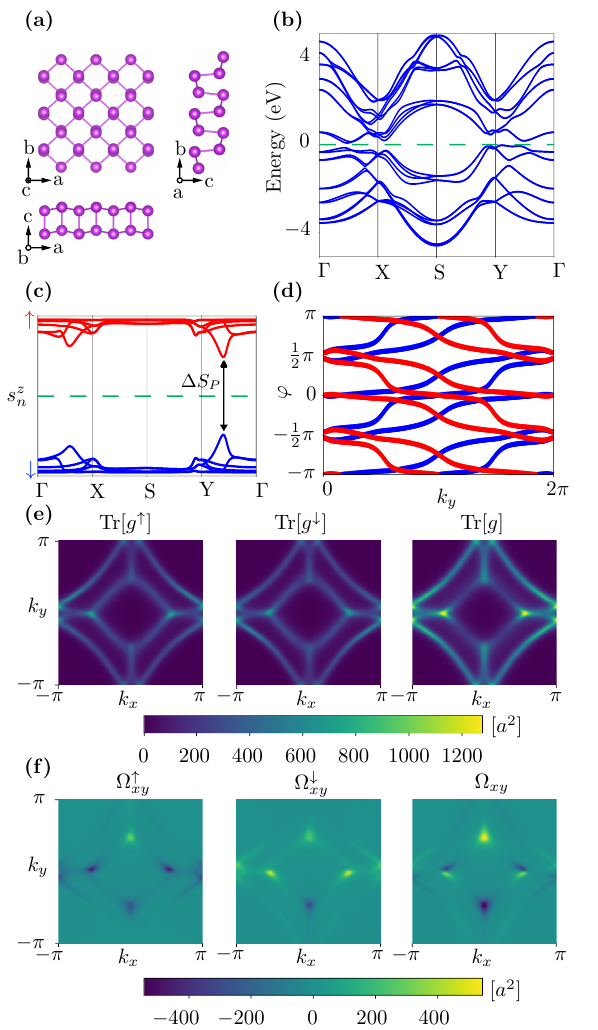}
    \caption{Quantum (spin-)geometry in ultrathin Bi. \textbf{(a)} Crystal structure, corresponding to the puckered $Pmn2_1$ phase \cite{UltrathinBiExperimental} of ultrathin Bi. \textbf{(b)} Band structure. \textbf{(c)} Spin-$z$ band structure with manifestly large spin gap $\Delta S_P$. \textbf{(d)} Topological winding of projected spin operator $PS_zP$ eigenstates in a spin-Wilson loop. \textbf{(e)} Metrics, where we observe that the quantum metric ($g_{ij}$) pattern is supported by the spin-resolved quantum metrics $g^\uparrow_{ij}$, $g^\downarrow_{ij}$. The spin-topological geometric bound is satisfied at every point within the BZ, with spin-Berry curvatures bounding the spin-resolved quantum metric from below \textbf{(f)}. We find as expected that $(2\pi)^{-1}\int_{\mathrm{BZ}}\Omega_{xy}^{\uparrow} = -(2\pi)^{-1}\int_{\mathrm{BZ}}\Omega_{xy}^{\downarrow} = 2$, whereas $(2\pi)^{-1}\int_{\mathrm{BZ}}\Omega_{xy}= 0.0$. By contrast, $(2\pi)^{-1}\int_{\mathrm{BZ}} \mathrm{Tr}[g^{\uparrow}]=(2\pi)^{-1}\int_{\mathrm{BZ}} \mathrm{Tr}[g^{\downarrow}] = 15.1$ whereas $(2\pi)^{-1}\int_{\mathrm{BZ}} \mathrm{Tr}[g] = 25.6$, demonstrating an enhanced quantum geometry in spin-topological Bi with $|C^\sigma_s| = 2$.} 
    \label{fig:FigBi}
\end{figure}

\begin{figure}[t!]
    \centering
    \def\svgwidth{0.95\linewidth}
    \includegraphics[width=\linewidth]{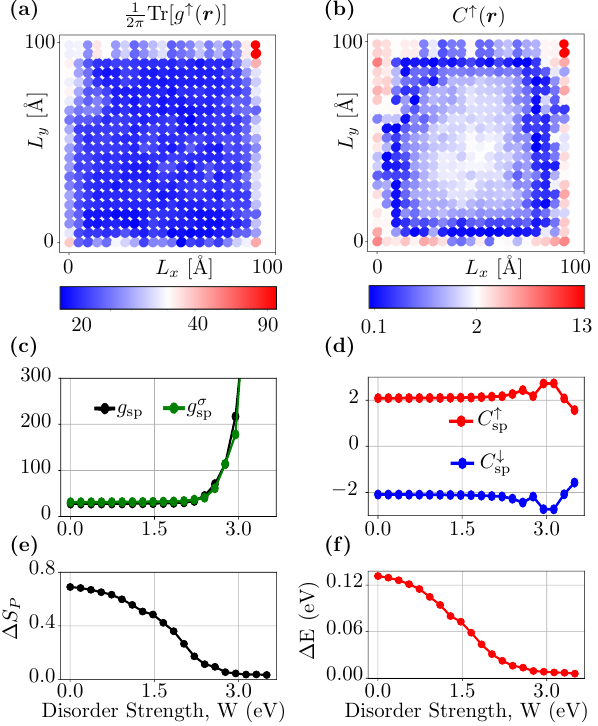}
    \caption{Robustness of spin-topologically bounded quantum geometry in the presence of disorder. \textbf{(a)} Spin-resolved metric markers $\frac{1}{2\pi} \text{Tr}[g^{\sigma}(\vec{r})]$ and \textbf{(b)} spin-Chern markers $C^\sigma(\vec{r})$ in disordered ultrathin Bi flake system of $1746$ atoms, with random potential disorder of strength ${W = 1.5~\text{eV}}$. Averaging over 120 atoms in the center yields: ${\frac{1}{2\pi} \overline{\text{Tr}~g^{\uparrow}(\vec{r})}=20.2}$ and ${\overline{C^{\uparrow}(\boldsymbol{r})} = 1.8}$. \textbf{(c)} Scaling of single-point metrics $g^{(\sigma)}_{\text{sp}} \equiv \frac{1}{2\pi} \text{Tr}[{g}^{(\sigma)}_{ij}]_\text{sp}$, \textbf{(d)} single-point spin-Chern numbers $C^\sigma_\text{sp}$, \textbf{(e)}~spin gap $\Delta S_p$, and \textbf{(f)} energy gap $\Delta $E against disorder strength $W$ in ultrathin Bi system of 784 atoms, averaged over 20 disorder realizations.} 
    \label{fig:FigDisorder}
\end{figure}

\sect{Quantum geometry in spin-topological Bi} We now demonstrate the interplay of spin-topologies with spin-resolved quantum geometries in the material context of ultrathin elemental bismuth. As a consequence, we further validate and confirm that the optical weights and the band gaps are topologically bounded. The associated first-principles calculations are detailed in Sec.~VI of the Supplemental Material~\cite{SI}.

In Fig.~\ref{fig:FigBi}, we consider ultrathin Bi in the puckered $Pmn2_1$ phase, as studied theoretically in Ref.~\cite{UltrathinBiPrediction} and experimentally in Ref.~\cite{UltrathinBiExperimental}. We find that it has spin-Chern numbers $|C_s| = 2$, as previously reported in Refs.~\cite{Peng2024,DoubledQSHBi}. As a consequence of the nontrivial spin topology, the spin-Wilson loop winds as shown in Fig.~\ref{fig:FigBi}(d). Having confirmed the nontrivial spin topology, we observe that the spin-Berry curvatures manifestly provide a local, i.e., holding at every point of the momentum space, bound on the spin-resolved quantum metrics [Figs.~\ref{fig:FigBi}(e)~and~\ref{fig:FigBi}(f)]. As a consequence, the total quantum metric realized by the electrons in the ground state of the material is lower-bounded by the $\mathbb{Z}$-index protected by the spin gap $\Delta S_P$ [Fig.~\ref{fig:FigBi}(c)], and we, moreover, find that the band gap of Bi $(E_g)$ satisfies the previously derived spin-topological quantum geometric bound. Intuitively, the large quantum metric in Bi, beyond the saturation of topological bound, likely arises in part due to positions of the orbitals within the unit cell \cite{Simon_geometry_dep}.

\begin{figure}[t!]
   \centering
   \def\svgwidth{0.95\linewidth}
    \includegraphics[width=\linewidth]{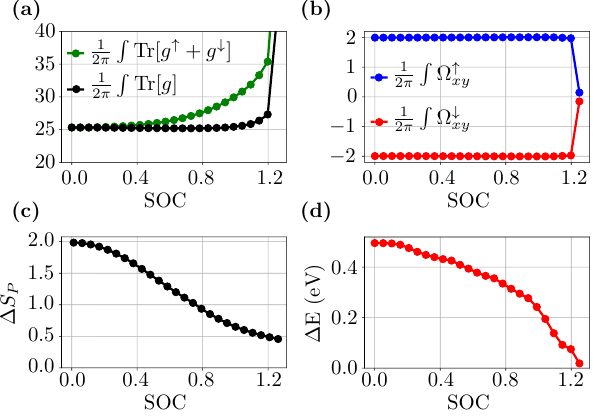}
    \caption{Scaling behavior of the quantum metric in ultrathin bismuth with SOC, where $\mathrm{SOC} = 1$ corresponds to the unperturbed Bi (for further information, see Sec.~VI of the Supplemental Material~\cite{SI}). We observe that \textbf{(a)} the metric diverges close to the critical value SOC $=1.2$, where the spin-Chern numbers \textbf{(b)} become ill defined. In this transition, the spin gap $\Delta S_P$ \textbf{(c)} remains well-defined up to the critical point, while the energy gap $\Delta E$ \textbf{(d)} closes. The analytical argument for the metric scaling is detailed in Sec.~VII of the Supplemental Material~\cite{SI}.} 
    \label{fig:scaling}
\end{figure}

\sect{Effects of disorder} We further validate the robustness of these results under the disorder present in realistic materials. In Fig.~\ref{fig:FigDisorder}, we demonstrate that the bounds are stable up to disorder-induced delocalization transitions. We confirm the robustness of geometric bounds for model Bi nanoflakes with an evaluation of real-space spin-geometric and spin-Chern markers equivalent to the momentum-space invariants under the change of basis~\cite{Bau2024}, see Figs.~\ref{fig:FigDisorder}(a)~and~\ref{fig:FigDisorder}(b), and single-point spin-Chern numbers~\cite{Favata2023} and metrics equal to the momentum-space definitions in a supercell within a single-point ($\vec{k}=0$) integral approximation~\cite{Ceresoli2007}, see Figs.~\ref{fig:FigDisorder}(c)~and~\ref{fig:FigDisorder}(d) and Appendixes A~and~B for technical details. Remarkably, we find that no additional symmetries are needed to protect the bound against the naturally present disorder, as long as the spin and energy gaps are preserved, see Figs.~\ref{fig:FigDisorder}(e)~and~\ref{fig:FigDisorder}(f), which is a promising indication that the retrieved geometric bounds hold in real materials under experimental conditions. Further details on the spin-geometric markers and disorder calculations are provided in Secs.~IX~and~X of the Supplemental Material~\cite{SI}, respectively.

\sect{Spin-resolved quantum metrology} We now discuss how the spin-resolved quantum geometry can be utilized for quantum metrology~\cite{Liu2020}, providing for a set of  applications within quantum sensing~\cite{Degen2017}, and opening a new direction of creating topological quantum sensors with time-reversal symmetric spin-topological materials, given their topologically bounded quantum geometry.  As such, an implementation of a spin-topological Hamiltonian in a quantum, e.g., free fermion, sensor~\cite{Sarkar2022}, offers for a topologically $\mathbb{Z}$-bounded enhancement of measurement sensitivity in realistic $\mathcal{T}$-symmetric, spinful systems. Quantum sensing relies on the uncertainty in the Cram\'er-Rao bounds~\cite{Degen2017}, which by relation to the quantum Fisher information~\cite{Liu2020} can be improved with enhanced quantum metric. We demonstrate the scaling of the quantum sensing-adaptable metric against the SOC realized in the spin-topological Bi Hamiltonian (Fig.~\ref{fig:scaling}), which is controllable with external parameters, e.g., bending~\cite{SOC_bending}, thus reducing uncertainty in the Cram\'er-Rao bounds for quantum sensing ({see also Appendix~C}).

\sect{Discussion} Below we discuss our results and their applicability. We first stress that the derivation of the here introduced spin-topological bounds was possible because of the presence of a spin gap, which defines the $P_\uparrow$, $P_\downarrow$ eigenstate manifolds. Closing the spin gap amounts to the trivialization of the here-derived geometric bounds, as well as of the spin-Chern numbers $C^\sigma_s$~\cite{Sheng2006, Prodan2009}. Moreover, the derivation of quantum geometric bounds is possible without a direct use of energy eigenstates; previously a similar strategy was implemented for deriving geometric bounds in topological Euler bands~\cite{Bouhon2023geometric, jankowski2023optical, Kwon2024}, where auxiliary states rather than eigenstates were utilized for the implementation of a geometric positive semidefiniteness condition, before ultimately connecting to the energy eigenstates via a unitary transformation. It should be noted that our bounds, rather than being of $\mathbb{Z}_2$ type, i.e., given by topological $\mathbb{Z}_2$ indices, as previously retrieved in electronic~\cite{Herzog2022, yu2024Z2} and excitonic contexts~\cite{jankowski2024excitons, thompson2024excitons}, fundamentally belong to the class of $\mathbb{Z}$ bounds~\cite{Rahul2014, Xie2020, Bouhon2023geometric, jankowski2023optical, Kwon2024}. Having derived new geometric optical bounds due to spin topology, a question of their experimental validation naturally arises. In that regard, quantum geometry was shown to be measurable with inelastic X-ray scattering~\cite{balut2024} and photoemission~\cite{Kang2024}, which also applies to the measurements of the quantum metric induced by spin topology studied in this work. A further experimental study of an interplay of the spin-resolved quantum geometry, spin-density response, and entanglement provides for an interesting future pursuit, in particular, given the robustness of our findings in the presence of ubiquitous disorders. 
Finally, we stress that while preserving a $\mathbb{Z}$ character, our bounds applicable to quantum metrology and quantum sensing remain nontrivial in the presence of time-reversal symmetry, unlike the previous bounds for topological magnetic systems~\cite{SciPostMera2022, Sarkar2022}. Hence, our bounds, adaptable even for four-state systems as discussed in the SM \cite{SI}, could enable a  class of topological quantum sensors operating without magnetic environments.

    \sect{Conclusions and Outlook} We propose an avenue for probing a range of spin-topological systems from the perspective of quantum geometric relations, which notably, for a vanishing $\mathbb{Z}_2$ index, were previously not targetable with observables. We derive a set of spin-topological $\mathbb{Z}$ bounds, even for eigenstates which are trivial from a Wilson loop perspective, which provide optical constraints on optical weights and bands gaps. We show that the bounds are valid in the presence of {disorder and} up to moderate spin-orbit entanglements, i.e., as long as the spin gap is preserved. We validated these quantum geometric conditions in bismuth and in an effective Hamiltonian with high spin-Chern number. By deriving  quantum geometric relations central to quantum metrology, we show that quantum geometry bounded by the spin topology could enable spin-topological quantum sensors.\\

    \sect{Acknowledgments} We thank B. Peng for useful discussions. W.J.J.~acknowledges funding from the Rod Smallwood Studentship at Trinity College, Cambridge. R.-J.S. acknowledges funding from an EPSRC ERC underwrite Grant No.  EP/X025829/1 and a Royal Society exchange Grant No. IES/R1/221060, as well as Trinity College, Cambridge. G. F. L. acknowledges funding from the European Union's Horizon Europe research and innovation programme under the Marie Skłodowska-Curie Grant Agreement No. 101126636. \\

\bibliography{references}

\section*{Appendices}

\sect{Appendix A: Spin-geometric and topological markers in disordered systems} We here provide minimal technical details on the spin-geometric real-space markers and their interplay with the spin-Chern markers. For full technical details, see Sec.~IX of the SM \cite{SI}.

The local spin~metric marker, which we define in real space under open boundary conditions for disordered spin-topological systems studied in this work, reads
\beq{}
g^\sigma_{ij}(\textbf{r}) = \frac{1}{2 A_\text{cell}} \sum_\alpha \bra{r_\alpha} P_\sigma \{P_\sigma \hat{x}_i P_\sigma, P_\sigma \hat{x}_j P_\sigma \} \ket{r_\alpha}.
\eeq
$\ket{r_\alpha}$ is a real-space physical orbital $\alpha$ at a unit cell centered at the position vector $\vec{r}$, $A_\text{cell}$ is the area of a unit cell, and $\{ \ldots, \ldots \}$ denotes an anticommutator of spin-projected operators $P_\sigma \hat{x}_i P_\sigma$, with spin-projector $P_\sigma$ in real-space basis and position operator components $\hat{x}_i$. The spin-Chern marker is defined as~\cite{Bau2024}
\beq{}
 C^\sigma (\textbf{r}) = \frac{4\pi}{A_\text{cell}} \mathfrak{Im}~ \sum_\alpha \bra{r_\alpha} P_\sigma [\hat{x}, P_\sigma] [\hat{y}, P_\sigma] \ket{r_\alpha}.
\eeq
On defining $\text{Tr}[g^\sigma(\textbf{r})] \equiv g_{xx}^\sigma(\textbf{r}) + g_{yy}^\sigma(\textbf{r})$, the spin~metric marker and spin-Chern marker satisfy a topological bound consistent with the results presented in Fig.~\ref{fig:FigDisorder},
\beq{eq:realspacebound}
    \sum_{i \in \text{cells}} \frac{1}{2\pi} \text{Tr}[g^\sigma(\textbf{r}_i)] \geq \sum_{i \in \text{cells}} |C^\sigma(\textbf{r}_i)|, 
\eeq
which we rigorously derive in Sec.~IX of the Supplemental Material. Supplementary calculations with spin~metric and spin-Chern markers are presented in Sec.~X of the Supplemental Material.

\sect{Appendix B: Single-point spin-geometric and spin-topological invariants in disordered systems} In the following, we outline minimal technical details on the single-point spin~metric and its interplay with the single-point spin-Chern invariant~\cite{Favata2023}. For full technical details, see Sec.~IX of the Supplemental Material. 

The single-point formulation relies on approximating the momentum-space invariants, as in Eq.~\eqref{eq:invbound}, with a single $k$-point sampling at a supercell BZ center ($\vec{k} = \vec{0}$), i.e., the $\Gamma$-point, assuming that the finite-size corrections of the shrunk BZ can be neglected~\cite{Ceresoli2007}. In the limit of infinite (disordered) supercell, the supercell BZ shrinks to the $\Gamma$-point, justifying the approximation. 

In this limit, the single-point spin-Chern numbers $C^\sigma_\text{sp}$ (${C^{\sigma}_s= \frac{1}{2\pi} \intBZ \dd \textbf{k}~ \text{Tr}~\Om^{\sigma}_{xy} \rightarrow C^\sigma_\text{sp}}$) are defined as~\cite{Favata2023}:
\beq{}
    C^\sigma_\text{sp} \equiv -\frac{|\vec{b}_{1}||\vec{b}_{2}|}{\pi} \sum^{N^\sigma_\text{occ}}_n~\mathfrak{Im}~ \langle \partial_{k_x} u_{n\boldsymbol{k}=\Gamma}^{\sigma}|\partial_{k_y} u_{n\boldsymbol{k}=\Gamma}^{\sigma}\rangle,
\eeq
where $\vec{b}_{1}$,~$\vec{b}_{2}$ are reciprocal lattice vectors of a two-dimensional supercell. Moreover, we here define the single-point (spin)~metric:
\beq{}
 [{g}^{(\sigma)}_{ij}]_\text{sp} \equiv 2|\vec{b}_{1}||\vec{b}_{2}|~ \sum^{N^{(\sigma)}_\text{occ}}_n~\mathfrak{Re}~ \langle \partial_{k_i} u_{n\boldsymbol{k}=\Gamma}^{(\sigma)}|\partial_{k_j} u_{n\boldsymbol{k}=\Gamma}^{(\sigma)}\rangle,
\eeq
with $\frac{1}{2\pi} \text{Tr}[{g}^{(\sigma)}_{ij}]_\text{sp} \equiv \frac{1}{2\pi} ([{g}^{(\sigma)}_{xx}]_\text{sp} + [{g}^{(\sigma)}_{yy}]_\text{sp})$. We retrieve a bound inherited from the local momentum-space inequality, Eq.~\eqref{eq:localbound}:
\beq{}
    \frac{1}{2\pi} \text{Tr}~[{g}^{\sigma}_{ij}]_\text{sp} 
     \geq |C^\sigma_\text{sp}|.
\eeq
Further details on the evaluation of numerical derivatives $|\partial_{k_i}u_{n\boldsymbol{k}=\Gamma}^{\sigma}\rangle$ are provided in Ref.~\cite{Favata2023}, and the further simplified expressions are detailed in Sec.~IX of the Supplemental Material~\cite{SI}.

\color{black}
\sect{Appendix C: Quantum Cram\'er-Rao bound and quantum sensing} The quantum metric, through its relations to the quantum Fisher information, sets a lower bound on multiparameter estimation with pure states, as captured by the so-called~quantum Cram\'er-Rao bound (QCRB), that is of central interest for quantum metrology and can be used in quantum sensing~\cite{Degen2017}. In the following, we consider pure states $\ket{\phi_\kv}$ with density matrix $\rho = \ket{\phi_\kv}\bra{\phi_\kv}$ constituted by the coherent superpositions of occupied single-particle eigenstates (see Sec.~VII of the Supplemental Material~\cite{SI} for more details). We furthermore provide a four-band model for $\mathbb{Z}$-indexed spin topology in Sec.~VIII of the Supplemental Material~\cite{SI}, which can be adapted for qudits or coupled qubits, and where the role of Bloch momenta can be taken by two angles, i.e.,~$(k_x,k_y) \rightarrow (\phi, \theta)$. The QCRB, given for a pure state by the quantum metric, reads~\cite{Liu2020}
\begin{equation}
	\sqrt{\text{det}~ \Sigma_{ij}} \geq  \frac{1}{4M \sqrt{\text{det}~g_{ij}}},
\end{equation}
where $M$ is the number of measurements and we introduce the covariance matrix $\Sigma_{ij}$ for an unbiased two-parameter estimator $\textbf{k}$. Here, the estimator that was identified with the angular parameters takes values in the two-parameter family over the two-torus [${\textbf{k} = (k_x, k_y) \in T^2}$] under a set of positive operator-valued measurements, ${\Pi_p }$ yielding the expected measurement values of individual parameters $\bar{k}_i$, with $i = x,y$. Explicitly, the covariance matrix reads \cite{Liu2020}
\begin{equation}
	\Sigma_{ij} = \langle \delta k_i \delta k_j \rangle \equiv \sum_{p} (k_i - \bar{k}_i) (k_j - \bar{k}_j) \text{Tr}[\rho \Pi_p] ,
\end{equation}
which quantifies the uncertainty in two measured parameters $k_i, k_j$. As we show numerically in spin-topological Hamiltonians (Fig.~\ref{fig:scaling}), $\Sigma_{ij}$ can be topologically suppressed by the metric values provided by our $\mathbb{Z}$~bound,  for spin-topological quantum sensors with topologically reduced uncertainties.

\end{document}

\title{SUPPLEMENTAL MATERIAL \\ Quantum geometric bounds in spinful systems with trivial band topology}

\author{Wojciech J. Jankowski}
\email{wjj25@cam.ac.uk}
%\thanks{}
\affiliation{TCM Group, Cavendish Laboratory, Department of Physics, J J Thomson Avenue, Cambridge CB3 0HE, United Kingdom}

\author{Robert-Jan Slager}
%\email{rjs269@cam.ac.uk}
\affiliation{Department of Physics and Astronomy, University of Manchester, Oxford Road, Manchester M13 9PL, United Kingdom}
\affiliation{TCM Group, Cavendish Laboratory, Department of Physics, J J Thomson Avenue, Cambridge CB3 0HE, United Kingdom}

\author{Gunnar F. Lange}
\email{g.f.lange@fys.uio.no}
%\thanks{}
\affiliation{Department of Physics, University of Oslo, N-0316 Oslo, Norway}
\affiliation{Centre for Materials Science and Nanotechnology, University of Oslo, N-0316 Oslo, Norway}

\date{\today}

\maketitle

\tableofcontents 

\newpage

\begin{widetext}
\section{Properties of the projected spin operator}\label{app:A}
In this Section, we summarize the relevant properties of the projected spin operator, $PS_{\hat{n}}P$, where $P(\kv) = \sum_n^{\text{occ}}\ket{u_{n\kv}}\bra{u_{n\kv}}$ projects onto an energetically isolated set of bands, and $S_{\hat{n}} = 1\otimes (\hat{n}\cdot\vec{\sigma})$ is the spin operator along $\hat{n}$ in the tight-binding basis with eigenvalues $\pm 1$. We focus on the projected spin-$z$ operator, $PS_zP$, in what follows. Additional discussions of projected spin operator can be found in \cite{Lin2024,Prodan2009,Lange2023}. We begin by noting that this operator can be written as:
\begin{equation}
    PS_zP = \sum_{n,m}^{N_\text{occ}} \bra{u_{n\kv}}S_z\ket{u_{m\kv}}\ket{u_{n\kv}}\bra{u_{m\kv}},
\end{equation}
from which it follows that the operator will always have $N-N_{\text{occ}}$ zero eigenvalues, associated with the unoccupied bands $Q=1-P$. We ignore these in the following, and focus on the remaining $N_{\text{occ}}$ eigenvalues $s_n^z(\kv)\in[-1,1]$, associated with the occupied states, which form the spin band structure. We note that these can be found by directly diagonalizing the reduced spin operator $[s_r]_{nm} = \bra{u_{n\kv}}S_z\ket{u_{m\kv}}$. We assume time-reversal symmetry (TRS)  throughout. TRS guarantees that $N_{\text{occ}}$ is even for any energetically isolated group of bands, implying that we can identify the projectors $P_{\uparrow} = \sum_{n}^{N_{\text{occ}}/2}\ket{u_{n\kv}^{\uparrow}}\bra{u_{n\kv}^{\uparrow}}$ and $P_{\downarrow}=  \sum_{n}^{N_{\text{occ}}/2}\ket{u_{n\kv}^{\downarrow}}\bra{u_{n\kv}^{\downarrow}}$ associated with the largest and smallest $N_{\text{occ}}/2$ projected spin eigenvalues $s_n^z$ respectively. Note that $P = P_{\uparrow}+P_{\downarrow}$, but that the states $\ket{u_{n\kv}^{\sigma}}$ are not in general energy eigenstates. We can further define the \textit{projected spin-gap} as:
\begin{equation}
\Delta S_P \equiv  \text{min}_{\kv} [s_{N_{\text{occ}}/2+1}^z(\kv)-s_{N_{\text{occ}}/2}^z(\kv)] \in [0,2].
\end{equation}
If spin is a good quantum number (i.e. $[S_z,H] = 0)$, then all energy eigenstates can be written as product states between orbital and spin degrees of freedom, so that $s_n^z = \pm 1$ and $\Delta S_P = 2$. We refer to this as the non spin-orbit entangled case. If $\Delta S_P < 2$, then we say that the system is \textit{spin-orbit entangled}. As such $\Delta S_P$ gives a rough estimate of the degree of entanglement between spin and orbital degrees of freedom within the energetically isolated band subspace spanned by $P(\kv)$. It was shown in Ref.~\cite{Lin2024} that time-reversal symmetry $\mathcal{T}$ guarantees $s_n^z(\boldsymbol{k}) = -s^z_{N_{\text{occ}}-n}(-\kv)$. Furthermore, inversion symmetry $\mathcal{P}$ guarantees that $s_n^z(\kv) = s_n^z(-\kv)$, which taken together implies that systems with $\mathcal{PT}$ symmetry satisfy $s_{n}^z(\kv) = -s_{N_{\text{occ}}-n}^z(\kv)$, implying that such systems have an effective chiral symmetry in the spin bands. We note finally that $\{\ket{u_{n\kv}}\}_{n=1}^{N_{\text{occ}}}$ furnish a complete basis for the space spanned by $P$, so that any eigenstate of $PS_zP$ in the image of $P$ (rather than the zero eigenstates associated with $Q= 1-P$) can be expanded as:
\begin{equation}\label{eq:coh}
    \ket{u_{m\kv}^{\sigma}} = \sum_n^{N_{\text{occ}}}c_{nm}^{\sigma}\ket{u_{n\kv}}.
\end{equation}
Manifestly, Eq.~\eqref{eq:coh} represents a $PS_zP$ operator eigenstate as a weighted coherent superposition of single-particle Hamiltonian eigenstates. In particular, in the absence of spin-orbit coupling (SOC), in the presence of spin conservation, we have $c_{nm}^{\sigma} = \delta_{nm}$, see Sec.~\ref{app::C} for further details.

\section{Spin-resolved quantum geometric bounds and their properties}\label{app::B}

Here we detail bounds arising from the spin-resolved quantum geometry introduced in the main text. We begin by writing the quantum geometric tensor (QGT) $Q_{ij}$ in terms of a projector onto occupied states $P$ at momentum $\textbf{k}$~\cite{provost1980riemannian}:
\beq{}
    Q_{ij} = \text{Tr}[P \partial_{k_i} P\partial_{k_j} P],
\eeq
as well as the spin-resolved quantum geometric tensor (spin-QGT),
\beq{}
    Q^\sigma_{ij} = \text{Tr}[P_\sigma \partial_{k_i} P_\sigma \partial_{k_j} P_\sigma],
\eeq
in terms of projectors $P_\sigma$ onto $P S_z P$ eigenstates, with $P_\uparrow + P_\downarrow = P$. On substituting the latter relation, we have:
\begin{eqnarray}
    Q_{ij} = Q^\uparrow_{ij}+Q^\downarrow_{ij}+\mathrm{Tr}[P_{\uparrow}\partial_iP_{\uparrow}\partial_j P_{\downarrow}] \nonumber \\
    +\mathrm{Tr}[P_{\uparrow}\partial_iP_{\downarrow}\partial_j P_{\uparrow}]+\mathrm{Tr}[P_{\uparrow}\partial_iP_{\downarrow}\partial_j P_{\downarrow}]\\
    +\mathrm{Tr}[P_{\downarrow}\partial_iP_{\uparrow}\partial_j P_{\uparrow}]+\mathrm{Tr}[P_{\downarrow}\partial_iP_{\downarrow}\partial_j P_{\uparrow}] \nonumber \\+\mathrm{Tr}[P_{\downarrow}\partial_iP_{\uparrow}\partial_j P_{\downarrow}] \nonumber,
\end{eqnarray}
where, as we show in the next Sections, the cross-terms vanish in the parallel-transport gauge, when the spin-orbit coupling (SOC) is vanishingly small.

Furthermore, beyond the projector-based spin-QGTs, we can define a non-Abelian spin-Berry connection matrix:
\beq{}
    A^{\sigma \sigma'}_{nm,i} = i\bra{u^\sigma_{n\kv}} \ket{\partial_{k_i} u^{\sigma'}_{m\kv}}, 
\eeq
which allows to recast the spin-resolved quantum metric, i.e., real part of the spin-QGT: $g^{\sigma}_{ij} \equiv\text{Re}~Q^\sigma_{ij}$: 
\beq{}
    g^\sigma_{ij} = \text{Re}~ \sum^{N^\sigma_\text{occ}}_n \left[ \bra{\partial_{k_i} u^\sigma_{n \kv}} \ket{\partial_{k_j} u^\sigma_{n \kv}} + A^{\sigma \sigma}_{nn,i} A^{\sigma \sigma}_{nn,j} \right].
\eeq
The imaginary part of spin-QGT gives diagonal elements of the spin-Berry curvature matrix:
\beq{}
    \text{Tr}~\Om^\sigma_{xy} = \sum^{N^\sigma_\text{occ}}_n \Om^\sigma_{xy,nn} = \sum^{N^\sigma_\text{occ}}_n i \left[ \bra{\partial_{k_x} u^\sigma_{n\kv}} \ket{\partial_{k_y}  u^\sigma_{n\kv}} - \text{c.c.}\right] = -2 \text{Im}~\sum^{N^\sigma_\text{occ}}_n~\bra{\partial_{k_x} u^{\sigma}_{n\kv}} (1-P_\sigma) \ket{\partial_{k_y} u^{\sigma}_{n\kv}} = -2 \text{Im}~Q^{\sigma}_{xy}.
\eeq
Centrally to this work, the spin-QGT is positive-semidefinite, which we prove below.

To begin the proof, we first decompose spin-QGT in terms of individual occupied spin-band contributions indexed with $n$, $Q^\sigma_{ij} = \sum^{N_\text{occ}}_n Q^{n,\sigma}_{ij}$:
\beq{}
    Q^{n,\sigma}_{ij} = \bra{\partial_{k_i} u^{\sigma}_{n\kv}} (1-P_\sigma) \ket{\partial_{k_j} u^{\sigma}_{n\kv}},
\eeq
with the projector yielding eigenvalues $\la_i = 0,1$, i.e., $P^\sigma_n \ket{u^{\sigma}_{m\kv}} = \delta_{mn} \ket{u^{\sigma}_{m\kv}}$. Hence, on constructing an auxiliary state $\ket{\phi} = \sum_i c_i \ket{\partial_{k_i} u^{\sigma}_{n\kv}}$ with arbitrary coefficients $c_i$, one obtains~\cite{Xie2020}
\beq{}
   \sum_{i,j} (c_i)^* Q^{n,\sigma}_{ij} c_j = \sum_{i,j} (c_i)^* \bra{\partial_{k_i} u^{\sigma}_{n\kv}} (1-P_\sigma) \ket{\partial_{k_j} u^{\sigma}_{n\kv}} (c_j) = \bra{\phi} (1-P_\sigma) \ket{\phi} \geq 0.
\eeq
Further on summing over all occupied spin-bands, we retrieve,
\beq{}
    \sum_{i,j} (c_i)^* Q^{\sigma}_{ij} c_j \geq 0,
\eeq
which shows the positive-semidefineteness of spin-QGT over the entire parameter space, analogously to the proofs of positive-semidefiniteness of QGTs of Refs.~\cite{Xie2020, Bouhon2023geometric}. As a consequence,
\beq{}
    \text{det}~Q^{\sigma}_{ij} = g^{\sigma}_{xx} g^{\sigma}_{yy} - (g^{\sigma}_{xy} - \frac{i}{2} \Om^{\sigma}_{xy}) (g^{\sigma}_{yx} - \frac{i}{2} \Om^{\sigma}_{yx}) = g^{\sigma}_{xx} g^{\sigma}_{yy} - g^{\sigma}_{xy} g^{\sigma}_{yx} - \frac{|\Om^{\sigma}_{xy}|^2}{4} \geq 0,
\eeq
where we used definitional symmetries $g^{\sigma}_{yx} = g^{\sigma}_{xy}$, $\Om^{\sigma}_{yx} = -\Om^{\sigma}_{xy}$. On rearranging and taking a square-root, we have:
\beq{}
\sqrt{\text{det}~g^{\sigma}_{ij}} \geq \frac{|\Om^{\sigma}_{xy}|}{2}, 
\eeq
and moreover, given that definitionally: $g_{xx}, g_{yy} >0$, and given that all metric components are real:
\beq{}
\text{Tr}~g^{\sigma}_{ij} \equiv g^{\sigma}_{xx} + g^{\sigma}_{yy} \geq 2\sqrt{g^{\sigma}_{xx}g^{\sigma}_{yy}} \geq 2 \sqrt{g^{\sigma}_{xx}g^{\sigma}_{yy}-(g^{\sigma}_{xy})^2} = \sqrt{\text{det}~g^{\sigma}_{ij}},
\eeq
where we used an inequality between an arithmetic and geometric means, as well as that the squares of real numbers are non-negative. On combining both here-derived inequalities, we retrieve the hierarchy of spin-resolved inequalities featured in the main text.

\section{Connecting spin-resolved quantum geometry to quantum geometry}\label{app::C}

We now derive the bound on the total (non spin-resolved) quantum metric due to the spin topology, which we further connect to the eigenstate transition dipole matrix elements in the optical contexts. We begin by 
writing the $P S_z P$ eigenstates as coherent superpositions of the occupied Hamiltonian eigenstates as in Eq.\,\eqref{eq:coh}. 
%
%\beq{}
%    \ket{u^\sigma_{m \kv}} = \sum^{N_\text{occ}}_n c^{\sigma}_{nm} \ket{u_{n \kv}} 
%\eeq
%
We substitute these coherent expansions to connect the matrix elements of the $PS_zP$ eigenstates and Hamiltonian eigenstates:

\beq{}
\begin{split}
\begin{aligned}
    \Big|\bra{u^\sigma_{m \kv}}  \ket{\partial_{k_i} u_{l \kv}}\Big|^2 &= \Big|\sum^{N_\text{occ}}_n (c^{\sigma}_{nm})^* \bra{u_{n \kv}} \ket{\partial_{k_i}u_{l \kv}} \Big|^2 \\&= \sum^{N_\text{occ}}_n |c^{\sigma}_{nm}|^2 |\bra{u_{n \kv}} \ket{\partial_{k_i}u_{l \kv}} |^2 + 2 \sum_{m' \neq n'} \text{Re} \Big[ (c^{\sigma}_{n'm})^*  c^{\sigma}_{m'm} \bra{u_{n' \kv}} \ket{\partial_{k_i}u_{l \kv}} \bra{u_{l \kv}} \ket{\partial_{k_i}u_{m' \kv}} \Big] \\ &\leq 2 \sum^{N_\text{occ}}_n |c^{\sigma}_{nm}|^2 |\bra{u_{n \kv}} \ket{\partial_{k_i}u_{l \kv}} |^2 \leq \Big(\sum^{N_\text{occ}}_n |c^{\sigma}_{nm}|^2 \Big) \Big(\sum^{N_\text{occ}}_n |\bra{u_{n \kv}} \ket{\partial_{k_i}u_{l \kv}} |^2\Big) = \sum^{N_\text{occ}}_n \Big|\bra{u_{n \kv}} \ket{\partial_{k_i}u_{l \kv}} \Big|^2,
\end{aligned}
\end{split}
\eeq
where we used the normalization condition within the occupied state manifold:
\beq{}
    \sum^{N_\text{occ}}_n |c^{\sigma}_{nm}|^2 = 1,
\eeq{}
which states that a $PS_zP$ operator eigenstate is a certain normalized superposition of the occupied Hamiltonian eigenstates with momentum $\textbf{k}$. More compactly, this inequality follows from the Cauchy-Schwarz inequality:
\beq{}
    \Big| \sum_n v^*_n w_n \Big|^2 \leq \Big(\sum_n |v_n |^2 \Big)\Big(\sum_n |w_n|^2\Big),
\eeq
where $v_n \equiv c^{\sigma}_{nm}$ and $w_n \equiv \bra{u_{n \kv}}  \ket{\partial_{k_i} u_{l \kv}}$ with a fixed derivative $ \partial_{k_i}$ and fixed eigenstate $\ket{u_{l \kv}}$.
Similarly, for matrix elements of type: 
\beq{}
\begin{split}
    \Big| \bra{u^\sigma_{m \kv}} \ket{\partial_{k_i}u^{\sigma'}_{l \kv}} \Big|^2 = \Big|\sum^{N_\text{occ}}_n (c^{\sigma}_{nm})^* \bra{u_{n \kv}} \ket{\partial_{k_i}u^{\sigma'}_{l \kv}} \Big|^2 \leq \sum^{N_\text{occ}}_{n} \Big|\bra{u_{n \kv}} \ket{\partial_{k_i}u^{\sigma'}_{l \kv}} \Big|^2 \\= \sum^{N_\text{occ}}_{n} \Big| \sum^{N_\text{occ}}_{n'} c^{\sigma}_{n'l}  \bra{u_{n \kv}} \ket{\partial_{k_i}u_{n' \kv}} \Big|^2 \leq \sum^{N_\text{occ}}_{n} \sum^{N_\text{occ}}_{n'} \Big| \bra{u_{n \kv}} \ket{\partial_{k_i}u_{n' \kv}} \Big|^2.
\end{split}
\eeq
From positive-semidefiniteness of $Q^{\sigma}_{ij}$, as derived in the previous Section:
\beq{eq:SpinBound}
    g^{\sigma}_{xx} + g^{\sigma}_{yy} \geq |\Om^{\sigma}_{xy}|.
\eeq
On combining both spin-up and spin-down inequalities:
\beq{}
    g^{\uparrow}_{xx} + g^{\uparrow}_{yy} + g^{\downarrow}_{xx} + g^{\downarrow}_{yy} \geq |\Om^{\uparrow}_{xy}| + |\Om^{\downarrow}_{xy}|,
\eeq
Given the diagonal elements of the spin-QGT are real by definition, we can recast the above inequality as:
\beq{}
     \sum^{N_\text{occ}/2}_m  \sum_{i=x,y} \sum_{\sigma= \uparrow, \downarrow} \bra{\partial_{i} u^{\sigma}_{m \kv}} (1- P_\sigma) \ket{\partial_{i} u^{\sigma}_{m \kv}} =   \sum^{N_\text{occ}/2}_m  \sum_{i=x,y} \sum_{\sigma= \uparrow, \downarrow} \bra{\partial_{i} u^{\sigma}_{m \kv}} (Q + P_{\sigma' \neq \sigma}) \ket{\partial_{i} u^{\sigma}_{m \kv}} \geq |\Om_{xy}^{\uparrow}| + |\Om_{xy}^{\downarrow}|,
\eeq
where we used the resolution of identity $1 = P_\uparrow + P_\downarrow + Q$, where $Q$ is a projector onto unoccupied bands. Splitting the resolved elements further, we obtain:
\beq{}
    \sum^{N_\text{occ}/2}_m \sum^{\text{unocc}}_{l}  \sum_{i=x,y} \sum_{\sigma= \uparrow, \downarrow} \Big|\bra{u^{\sigma}_{m \kv}} \ket{\partial_{k_i} u_{l \kv}}\Big|^2 + \sum^{N_\text{occ}/2}_{m,m'}  \sum_{i=x,y} \sum_{\sigma= \uparrow, \downarrow} \sum_{\sigma' \neq \sigma} \Big|\bra{u^{\sigma}_{m \kv}} \ket{\partial_{k_i} u^{\sigma'}_{m' \kv}}\Big|^2 \geq |\Om_{xy}^{\uparrow}| + |\Om_{xy}^{\downarrow}|,
\eeq
and on employing the inequalities for individual matrix elements derived above,
\beq{}
    \sum^{N_\text{occ}/2}_{m}\sum^{N_{\text{occ}}}_{n} \sum^{\text{unocc}}_{l} \sum_{i=x,y} \sum_{\sigma= \uparrow, \downarrow} |\bra{u_{n \kv}} \ket{\partial_{k_i} u_{l \kv}}|^2 + \sum^{N_\text{occ}/2}_{m,m'} \sum^{N_\text{occ}}_{n,n'}  \sum_{i=x,y} \sum_{\sigma= \uparrow, \downarrow} |\bra{u_{n \kv}} \ket{\partial_{k_i} u_{n' \kv}}|^2  \geq |\Om_{xy}^{\uparrow}| + |\Om_{xy}^{\downarrow}|.
\eeq
On performing the redundant summations:
\beq{}
     2(N_{\text{occ}}/2) \sum^{N_\text{occ}}_{n} \sum^{\text{unocc}}_{l} \sum_{i=x,y} |\bra{u_{n \kv}} \ket{\partial_{k_i} u_{l \kv}}|^2 + 2(N^2_{\text{occ}}/4) \sum^{N_\text{occ}}_{n,n'}  \sum_{i=x,y}  |\bra{u_{n \kv}} \ket{\partial_{k_i} u_{n' \kv}}|^2 \geq |\Om_{xy}^{\uparrow}| + |\Om_{xy}^{\downarrow}|
\eeq
\beq{}
    N_{\text{occ}} \sum_{i=x,y} g_{ii} +(N^2_{\text{occ}}/2) \sum^{N_\text{occ}}_{n,n'}  \sum_{i=x,y} g^{nn'}_{ii}  \geq |\Om_{xy}^{\uparrow}| + |\Om_{xy}^{\downarrow}|.
\eeq
Notably, in a parallel-transport gauge, the multiband matrix elements \textit{between the occupied bands} $g^{nn'}_{ii}$ vanish, unlike the interband matrix elements between occupied and unoccupied bands $\sum^{N_\text{occ}}_{n} \sum^{\text{unocc}}_{l} g^{nl}_{ii} = g^{}_{ii}$. To demonstrate this, we note that the non-Abelian Berry connection in the occupied bands is a Hemitian matrix, i.e., $\vec{A}_{nn'} = (\vec{A}^*_{n'n})$, and hence, it can be diagonalized in the occupied band subspace by a similarity transformation: $\vec{A}_{nn'} \rightarrow U\vec{A}_{nn'} U^\dagger$, with $U$ the matrix of eigenvectors of $\vec{A}_{nn'}$. The unitary change of basis within the occupied band subspace corresponds to choosing the parallel-transport gauge. Hence, we have,
\beq{}
    g^{nn'}_{ij} \equiv \bra{\partial_{k_i} u_{n \kv}} \ket{ u_{n' \kv}} \bra{u_{n' \kv}} \ket{\partial_{k_i} u_{n \kv}} = A^i_{n n'} A^j_{n' n} = \delta_{n n'} g^{nn}_{ij}. 
\eeq
Finally, we recognize that in the context of this work, the leftover terms $g^{nn}_{ii} = |A^i_{n n}|^2$ can be locally gauged away by the $U(1)$ gauge transformation: $\ket{u_{n\kv}} \rightarrow e^{i\alpha(\kv)} \ket{u_{n\kv}}$, resulting in $A^i_{n n} \rightarrow A^i_{n n} + \partial_{k_i} \alpha(\kv) = 0$, on fixing a smooth gauge \textit{locally}. We note that in both the systems with trivial or nontrivial Wilson loop windings [independently from the nontrivial spin Wilson loop], such local gauge fixing amounts to a local bound at a given $\textbf{k}$-point, which is equivalent to fixing the gauge to a local trivialization with $||\partial_{k_i} P_{n\kv}|| \equiv \sqrt{\text{Tr} (\partial_{k_i} P_{n\kv})^2} = 0$, where $P_{n\kv} = \ket{u_{n\kv}} \bra{u_{n\kv}}$ is a local projector. Hence, on applying a parallel-transport gauge, we arrive at a local bound,
\beq{eq:CentralBound}
    N_{\text{occ}} \sum_{i=x,y} g_{ii} \geq |\Om_{xy}^{\uparrow}| + |\Om_{xy}^{\downarrow}|,
\eeq
in a general spin-orbit coupled, and spin non-conserving, case. This concludes the derivation of the fundamental quantum geometric bound on connecting the spin-resolved quantum geometry to a standard multiband quantum geometry~\cite{Ahn2020, Ahn2021, Bouhon2023geometric}.

Finally, as a special case, in the spin-conserving case, we retrieve a stronger inequality, consistently with Ref.~\cite{Lysne2023}. Namely, the absence of spin-mixing SOC, allows to sharpen the previous inequality,
\beq{}
\begin{split}
    \Big| \bra{u^\sigma_{m \kv}} \ket{\partial_{k_i}u^{}_{l \kv}} \Big|^2 =  \sum^{N_\text{occ}}_{n} \Big| \bra{u_{n \kv}} \ket{\partial_{k_i}u_{l \kv}} \Big|^2 \rightarrow \Big| \bra{u_{m \kv}} \ket{\partial_{k_i}u_{l \kv}} \Big|^2,
\end{split}
\eeq
as the sum, $\sum_n |c^\sigma_{nm}|^2 = |c^\sigma_{mm}|^2 = 1$, reduces to a single term. On repeating the previous steps with reduction, one instead obtains a local bound,
\beq{}
    \sum_{i=x,y} g_{ii} \geq |\Om_{xy}^{\uparrow}| + |\Om_{xy}^{\downarrow}|,
\eeq
which on integrating one obtains the bound due to the spin-Chern numbers~\cite{Lysne2023}. We note that both the bounds in the spin-conserving and non-conserving cases also hold in systems without inversion symmetries, which were \textit{not} assumed in the derivations above. 

\section{Effects of spin-orbit coupling and spin-orbit entanglement}\label{app::D}
We introduce spin-orbit coupling as a perturbation $\Delta H = \sum_{n,m} \lambda_{nm} (1-\delta_{\sigma \sigma'}) \ket{\psi^{\sigma}_{n\kv}}\bra{\psi^{\sigma'}_{m\kv}}$, where the limit $\lambda = 0$ corresponds to the vanishingly small SOC.
The time-independent perturbation theory applied to first order, yields:
\beq{}
    \ket{\psi_{n\kv}}=\ket{\psi^{\lambda=0}_{n\kv}} + \sum_{m \neq n} \frac{\bra{\psi^{\lambda=0}_{m\kv}} \Delta H \ket{\psi^{\lambda=0}_{n\kv}}}{E_{n\kv}-E_{m\kv}} \ket{\psi^{\lambda=0}_{m\kv}},
\eeq
where $E_{n\kv}$ are the energy eigenvalues of the non-degenerate eigenstates for $\lambda = 0$. Here, we use the basis of Bloch states ${\ket{\psi_{m\kv}} = e^{i\kv \cdot \vec{\hat{r}}} \ket{u_{m\kv}}}$. Moreover, when $\lambda = 0$, the Hamiltonian eigenstates are also the eigenstates of the $PS_zP$ operator, $\ket{\psi^{\lambda =0}_{n\kv}} = \ket{\psi^\sigma_{n\kv}}$, thus, we have:
\beq{}
    \ket{\psi_{n\kv}}=\ket{\psi^\sigma_{n\kv}} + \sum_{m \neq n} \frac{\lambda_{mn} (1-\delta_{\sigma \sigma'}) }{E_{n\kv}-E_{m\kv}} \ket{\psi^{\sigma'}_{m\kv}}.
\eeq
In the following, we further demonstrate the scaling of the introduced metrics with the spin-orbit entanglement. For simplicity, we set $\lambda_{mn} = \lambda$ and assume a flat-band dispersion $E_{n\kv} = E_n$. On substituting the perturbative correction, we begin by writing the spin-resolved quantum metrics:
\beq{}
    g^\sigma_{ij}(\lambda) = g^\sigma_{ij}(\lambda = 0)  - \sum_{m \neq n}  \frac{\lambda}{E_n - E_m} A^{\sigma \sigma'}_{nm, i} A^{\sigma' \sigma}_{nm, j} + \sum_{m \neq n}  \frac{\lambda^2}{(E_n - E_m)^2} A^{\sigma \sigma'}_{nm, i} A^{\sigma' \sigma}_{mn, j}+ O(\lambda^3),
\eeq
which at the leading order allows for a possible reduction of the $g_{ij}$ in the eigenstate basis, while satisfying the lower bound with an addition of the spin-entanglement factor $1/N_{\text{occ}}$, as demonstrated in Sec.~IV. Numerically, we find that instead the metrics increase in magnitude with the increasing $\lambda$, as shown in the main text, which is possible because the band gap and spin gaps reduce in size, as the SOC is increased. While in the above we treat $\la$ and $|E_n - E_m|$ as independent parameters, these are indeed not independent for any local Hamiltonian, which explains why the numerically observed scaling of the metric is increasing rather than decreasing with SOC.

\section{Spin-resolved quantum geometry and optical responses}\label{app::E}

As in the previous Sections, we begin by noting that $P_{\uparrow} = P-P_{\downarrow}$, so that $1-P_{\uparrow} = 1-P+P_{\downarrow}$. Therefore, intuitively, one can interpret $Q_{ij}^{\uparrow}$ as transitions into either the unoccupied manifold \textit{or} the occupied states corresponding to projected spin-down states from projected spin-up states. 

To connect to optics, we make the following identification, following Refs.~\cite{Bianco2011, Ahn2021},
\beq{}
    \mathcal{M} \equiv |\bra{u_{n\kv}}\ket{\partial_{k_i} u_{m\kv}}|^2 = |\bra{\psi_{n\kv}} \hat{r}_i \ket{\psi_{m\kv}}|^2
\eeq
namely that the non-Abelian Berry connection elements can be rewritten in terms of transition dipole matrix elements for E1 optical transitions, induced by the electron-photon coupling that reads $\Delta H = e\vec{E}(t) \cdot\vec{ \hat{r}}$ in the length gauge, with $\vec{\hat
r}$ the position oeprator. As such, we can connect the matrix elements to the optical transition rates using Fermi's golden rule:
\beq{}
    \Gamma_{i} (\om) = \frac{2\pi}{\hbar} \sum^{N_\text{occ}}_n \sum^{\text{unocc}}_m |\bra{\psi_{n\kv}} \hat{r}_i \ket{\psi_{m\kv}}|^2 \delta (E_{m\kv} - E_{n\kv} - \hbar \omega), 
\eeq
which we connect to quantum metric~\cite{Ozawa2018}, to retrieve a new bound
\beq{}
    \Gamma_x + \Gamma_y = \frac{2\pi e^2 |\vec{E}|^2}{\hbar^2} \intBZ \dd^2 \kv~(g_{xx} + g_{yy}) \geq \frac{2\pi e^2 |\vec{E}|^2}{N_{\text{occ}}\hbar^2} (|C^{\uparrow}_s| + |C^{\downarrow}_s|),
\eeq
with $\Gamma_{x/y} \equiv \int^\infty_0 \dd \om~\Gamma_{x/y} (\om)$ denoting integrated optical transition rates on coupling to $x$- and $y$-polarized light. Here, we assumed that spin is not necessarily conserved. In the presence of spin conservation, we analogously have:
\beq{}
    \Gamma_x + \Gamma_y \geq \frac{2\pi e^2 |\vec{E}|^2}{\hbar^2} (|C^{\uparrow}_s| + |C^{\downarrow}_s|).
\eeq
Further to the bounds on transition rates, we connect the geometric bound to optical conductivity $\sigma_{ij}(\om)$. The optical conductivity reads
\beq{}
    \sigma_{ii}(\om) = i e^2 \intBZ \frac{\dd^2 \kv}{(2\pi)^2} \sum_{m,n}  (f_{n\kv} - f_{m \kv}) \frac{\hbar \om \bra{\psi_{n\kv}} \hat{v}_i \ket{\psi_{m\kv}} \bra{\psi_{m\kv}} \hat{v}_i \ket{\psi_{n\kv}}}{(E_{n\kv}-E_{m\kv})^2 (\om-E_{m\kv}/\hbar + E_{n\kv}/\hbar + i0^+)},
\eeq
where $\hat{v}_i = \frac{i}{\hbar} [\hat{H}, \hat{r_i}]$ is a velocity operator, and  $f_{n\kv}= \Big[\text{exp}\big((E_{n\kv}-\mu)/T\big)+1\Big]^{-1}$ with chemical potential $\mu$ are Fermi-Dirac thermal occupation factors, which in the zero temperature $(T=0)$ limit $(f_{nm} = f_n - f_m = 1)$ reduces to~\cite{Ahn2021}
\beq{eq:multiCond}
    \sigma_{ii}(\om) =  \frac{\pi \om e^2}{\hbar} \intBZ \frac{\dd^2 \kv}{(2\pi)^2} \sum_{m,n} f_{nm} g^{mn}_{ii} \delta (\om - E_{m\kv}/\hbar + E_{n\kv}/\hbar).
\eeq
On integrating, one obtains a well-known relation between the optical conductivity and geometry, namely, the Souza-Wilkens-Martin (SWM) sum rule~\cite{Souza2000}
\beq{}
    \int^\infty_0 \dd \om~\frac{\text{Re}~\sigma_{ij}(\om)}{\om} = \frac{e^2}{2h} \intBZ \dd^2 \kv~g_{ij},
\eeq
which in the main text, we employ combined with the bound by the spin-Chern numbers. Moreover, to arrive at the band gap bound, we combine this result with the $f$-sum rule~\cite{Kivelson1982, Fu2024}:
\beq{}
    \frac{\pi}{2} \frac{n e^2}{m} = \int^\infty_0 \dd \om~\text{Re}~\sigma_{ii}(\om),
\eeq
which culminates in a hierarchy of inequalities,
\beq{}
    \frac{n h e^2}{2 m E_g} = \sum_{i=x,y} \frac{\hbar}{E_g} \int^\infty_0 \dd \om~\text{Re}~\sigma_{ii}(\om) \geq \sum_{i=x,y} \int^\infty_0 \dd \om~\frac{\text{Re}~\sigma_{ii}(\om)}{\om} = \sum_{i=x,y} \frac{e^2}{2h} \intBZ \dd^2 \kv~g_{ii} \geq \frac{e^2}{2\hbar}  S_\lambda.
\eeq
Here, we also assumed that the are no sub-gap excitations contributing to the optical conductivity $\sigma_{ii}(\om)$ at frequencies $0 < \om < E_g/\hbar$, which holds for single-particle states in the absence of many-body effects. Fig.\,\ref{fig:Optical_response} shows the optical response for ultrathin Bi, as well as the four-band TRS-symmetric QWZ model discussed in Sec.~VIII, where we define:
\begin{equation}\label{eq:def_optical_response}
    K(\omega_c) = \frac{2\hbar}{e^2}\int_0^{\omega_c} \dd\omega \frac{\mathrm{Re}[\sigma_{xx}(\omega)+\sigma_{yy}(\omega)]}{\omega} =\frac{1}{2\pi}\int_0^{\omega_c}\dd\omega \int_{\mathrm{BZ}} \dd^2\kv \sum_{mn}(g_{xx}^{mn}+g_{yy}^{mn})f_{nm}\delta(\hbar \omega - E_m+E_n),
\end{equation}
where as expected, in the zero temperature limit, $K(\omega_c\rightarrow \infty) = (2\pi)^{-1}\int_{\mathrm{BZ}} \mathrm{Tr}[g]$.

\begin{figure}
    \centering
    \def\svgwidth{0.75\linewidth}
    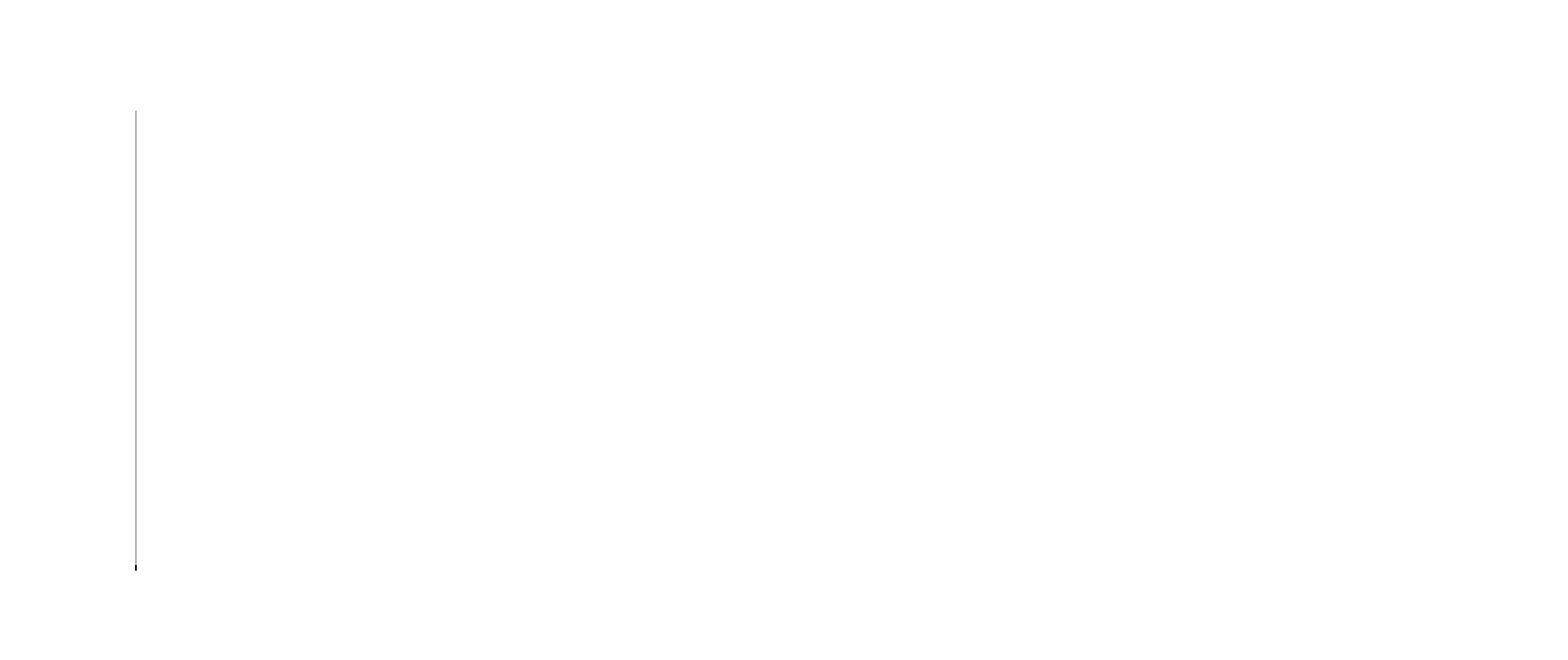
    \caption{Optical response for ultrathin Bi and the TRS-QWZ model in Eq.\,\eqref{eq:minimal_model}, with $K(\omega_c)$ defined in Eq.\,\eqref{eq:def_optical_response}. The red dotted line corresponds to the integral of the trace of the metric over the BZ, $(2\pi)^{-1}\int_{\mathrm{BZ}}\mathrm{Tr}[g]$.}    \label{fig:Optical_response}
\end{figure}

\section{First-principles calculations}\label{app::F}
The ultrathin bismuth band calculations are based on the computations obtained in Ref.\,\cite{Peng2024}. The calculations were performed using \texttt{QuantumEspresso} \cite{QE_1,QE_2}, with a fully relativistic PAW pseudopotential using the PBEsol functional. The k-point grid was chosen as 12$\times$12$\times$1, with a vacuum spacing of 24 Å, and a truncated Coloumb interaction in the $z$-direction \cite{ColoulmbTrunctation}, using a kinetic energy cutoff of 120 for the wavefunction and 500 Ry for the charge density. The output is then projected to maximally localized Wannier functions using \texttt{Wannier90} \cite{Wannier90_1,Wannier90_2,Wannier_90_3} for the $6p$ orbitals of bismuth. The resulting tight-binding models are then used for all subsequent calculations.

To get the SOC scaling in Fig.~3 of the main text, we write the tight-binding Hamiltonian in the basis $\{\ket{6p}\otimes \ket{\uparrow \downarrow}\}$, and uniformly scale all matrix elements that couple different spins with a single parameter $\lambda$, so that $\lambda = 1$ corresponds to the original model.

We further utilize the computed eigenstates to retrieve the optical conductivity $\sigma_{ij}(\om)$ and demonstrate the spin-topological bound realized by $\text{Re}~\sigma_{ij}(\om)$. To compute the optical response, we employ the gauge-invariant projectors onto occupied $P_{n \kv} = \sum_{E_n = E_{n_i}} \ket{u_{n_i\kv}} \bra{u_{n_i\kv}}$ and unoccupied bands $P_{m \kv} = \sum_{E_m = E_{m_j}} \ket{u_{m_j\kv}} \bra{u_{m_j\kv}}$~\cite{Ahn2021}, where $n = \{ n_i \}$ and $m = \{ m_j \}$ denote the sets of degenerate bands at momentum $\kv$ with energies $E_n$ and $E_m$, respectively. We employ these projectors to compute the multiband quantum metric contributions~\cite{Ahn2021},
\beq{}
    g_{ij} = \sum^{N_\text{occ}}_{n} \text{Re}~\bra{\partial_{k_i} u_{n\kv}} \Big( \sum^{\text{unocc}}_{m} 
 \ket{u_{m\kv}} \bra{u_{m\kv}} \Big) \ket{\partial_{k_j} u_{n\kv}} = \sum^{N_\text{occ}}_{n} \sum^{\text{unocc}}_{m} g^{mn}_{ij}  = \sum^{N_\text{occ}}_{n} \sum^{\text{unocc}}_{m} \text{Re}~ \bra{u_{m\kv}} \ket{\partial_{k_j} u_{n\kv}} \bra{\partial_{k_i} u_{n\kv}} \ket{u_{m\kv}},
\eeq
We further note that: $0 = \partial_{k_i} (\bra{u_{n\kv}} \ket{u_{m\kv}}) = \bra{\partial_{k_i} u_{n\kv}} \ket{u_{m\kv}} + \bra{ u_{n\kv}} \ket{\partial_{k_i}u_{m\kv}}$. By using (i) the product rule, (ii) orthogonality of eigenstates, (iii) cyclic property of trace, we further have: 
\beq{}
\begin{split}
    g^{mn}_{ij} = -\text{Re}~ \bra{u_{m\kv}} \ket{\partial_{k_j} u_{n\kv}} \bra{u_{n\kv}} \ket{\partial_{k_i} u_{m\kv}} = -\frac{1}{2} \bra{u_{m\kv}} \ket{\partial_{k_j} u_{n\kv}} \bra{u_{n\kv}} \ket{\partial_{k_i} u_{m\kv}} - \frac{1}{2}  \bra{\partial_{k_j} u_{n\kv}} \ket{u_{m\kv}} \bra{\partial_{k_i} u_{m\kv}} \ket{u_{n\kv}}  \\= -\frac{1}{2} \text{Tr} [(\ket{\partial_{k_j} u_{n\kv}} \bra{u_{n\kv}}) (\ket{\partial_{k_i} u_{m\kv}} \bra{u_{m\kv}} )] -\frac{1}{2} \text{Tr} [(\ket{ u_{n\kv}} \bra{\partial_{k_j} u_{n\kv}}) (\ket{u_{m\kv}} \bra{\partial_{k_i} u_{m\kv}} )] = -\frac{1}{2} \text{Tr} [\partial_{k_j} P_{n\kv} \partial_{k_i} P_{m\kv}],
\end{split}
\eeq
which on combining with the first-principles band energies and Eq.~\eqref{eq:multiCond} obtains the interband optical conductivity of bismuth in an insulating phase. We show our findings in Fig.~\ref{fig:Optical_response}, and observe that in the high-frequency limit, the optical weight saturates to the integral of the quantum metric computed in the main text.

\section{Spin-resolved quantum geometry and quantum metrology}\label{app::G}

In the following, we provide further details on spin-topological quantum metrology. Furthermore, we provide an analytical argument for the scaling of quantum Fisher information (QFI) with the spin-orbit entanglement, which is central to the main text, and where we retrieved this result numerically.

The quantum Fisher information, in the context of multiparameter quantum sensing~\cite{Degen2017} with states given by a density matrix $\rho = \sum^N_{n=1} \lambda_{n} \ket{\psi_{n}} \bra{\psi_{n}}$, where $\kv$ is a vector of parameters, can be defined as~\cite{Liu2020}:
\beq{}
    F_{ij}[\rho] \equiv \sum^{N}_{n,m} \frac{2 \text{Re}~(\bra{\psi_n} \partial_{k_i} \rho \ket{\psi_m} \bra{\psi_m} \partial_{k_j} \rho \ket{\psi_n}) }{\lambda_n + \lambda_m} 
\eeq
which intuitively captures how the probability distribution given by a density matrix is changed subject to changing a parameter associated with, i.e., conjugate to, an observable. Indeed, on recognizing the probability distributions associated with the density matrix and POVMs, one can obtain the classical Fisher information in the classical limit,
\beq{}
    F_{ij} = \sum_n \frac{1}{P(X_n,r)} \left( \frac{\partial P(X_n,r)}{\partial r} \right)^2,
\eeq
which reflects the overlaps and changes in the probability distribution $P(X_n,r)$ for a variable $r$, with outcomes $X_n$, subject to the parameter changes.

The quantum metric, here bounded by spin topology, fully determines the quantum Fisher information (QFI) in the case of \textit{pure} states. We consider a pure state $\ket{\phi_\kv} = \frac{1}{\sqrt{N_\text{occ}}} \sum^{N_\text{occ}}_{n=1} \ket{\psi_{n \kv}}$ with density matrix $\rho = \ket{\phi_\kv} \bra{\phi_\kv}$. The QFI matrix $F_{ij}[\rho]$ for a pure state, with $i,j$ parameter indices and with position operator $\hat{r}_i$ as an observable, reads~\cite{Liu2020}:
\begin{equation}
\begin{split}
	F_{ij}[\rho] = \sum_{m,n: \lambda_m + \lambda_n \geq 0} 2\frac{(\lambda_m - \lambda_n)^2}{\lambda_m + \lambda_n} \\ \bra{\psi_{m\kv}} \hat{r}_i \ket{\psi_{n\kv}}  \bra{\psi_{n\kv}} \hat{r}_j \ket{\psi_{m\kv}} = 4 g_{ij}.
\end{split}
\end{equation}
where $\lambda_m$ are the eigenvalues of the density matrix $\rho$, and which we recast in terms of quantum metric~\cite{Liu2020}. Correspondingly, the quantum Cram\'er-Rao bound (QCRB) is determined by QFI~\cite{Liu2020},
\begin{equation}
	\sqrt{\text{det}~ \Sigma_{ij}(\hat{\vec{k}})} \geq \frac{1}{M \sqrt{F_{ij}[\rho]}} = \frac{1}{4M \sqrt{\text{det}~g_{ij}(\textbf{k})}},
\end{equation}
where $\Sigma_{ij}(\hat{\vec{k}})$ is a covariance matrix for a two-parameter estimator $\hat{\vec{k}}$, as defined in the main text. The QCRB is central to the quantum-metrological applications considered and discussed in the main text.

We further consider a decomposition the density matrix in terms of projected spin contributions,
\beq{}
    \rho = \rho_\uparrow + \rho_\downarrow,
\eeq
which allows us to define a spin-projected quantum Fisher information (spin-QFI). The spin-QFI, for a spin-eigenstate, reads:
\beq{}
    F_{ij}[\rho_\sigma] = \sum_{m,n: \lambda_m + \lambda_n \geq 0} 2\frac{(\lambda_m - \lambda_n)^2}{\lambda_m + \lambda_n} \bra{\psi^{\sigma}_{m\kv}} \hat{r}_i \ket{\psi^{\sigma}_{n\kv}}  \bra{\psi^{\sigma}_{n\kv}} \hat{r}_j \ket{\psi^{\sigma}_{m\kv}} = 4g^\sigma_{ij},
\eeq
where in the last line, we connect the spin-resolved quantum geometry to the introduced estimator analogously to the standard QFI~\cite{Liu2020}.

Furthermore, we can consider a pure many-body fermionic state resolvable in single-particle quasiparticle states $\ket{\psi_{n\kv}}$ with a density matrix $\rho = \ket{\psi_\lambda} \bra{\psi_\lambda}$, which determines a many-body Fisher information~\cite{Sarkar2022}, 
\beq{}
     F^{\text{MB}}_{ij}[\rho] = 4\text{Re} \bra{\partial_{\lambda_i} \psi_\lambda} \ket{\partial_{\lambda_j} \psi_\lambda} - \bra{\partial_{\lambda_i} \psi_\lambda} \ket{\psi_\lambda} \bra{\psi_\lambda} \ket{\partial_{\lambda_j} \psi_\lambda},
\eeq
where $\lambda$ can be a SOC strength parameter that can be controlled with external environment. In particular, the Fisher information diverges at a critical SOC corresponding to a topological phase transition point $\la = \la_c$ due to the divergence of the metric, as previously retrieved for free fermion topological quantum sensors~\cite{Sarkar2022}. 

Alternatively, the parameter space $\{ \la \}$ can be spanned by the centre-of mass momenta $\vec{\la} = (K_x, K_y)$ of a many-body state. Correspondingly, we have
\begin{equation}
	\sqrt{\text{det}~ \Sigma_{ij}(\hat{\vec{K}})} \geq \frac{1}{M \sqrt{F^{\text{MB}}_{ij}[\rho]}} = \frac{1}{4M \sqrt{\text{det}~g^{\text{MB}}_{ij}(\textbf{K})}},
\end{equation}
where $\Sigma_{ij}(\hat{\vec{K}})$ is a covariance matrix for the two-parameter estimation using a many-body state $\ket{\psi_\vec{K}}$ and $g^{\text{MB}}_{ij} (\textbf{K})$ is a many-body metric,
\beq{}
    g^{\text{MB}}_{ij} (\textbf{K}) = \frac{1}{2} \bra{\partial_{K_i} \psi_\textbf{K}} (1 - \ket{\psi_\textbf{K}} \bra{\psi_\textbf{K}}) \ket{\partial_{K_j} \psi_\textbf{K}} + \text{c.c.}.
\eeq
On using $\vec{K} = 0$ and a Slater determinant many-body fermionic state $\ket{\psi_\textbf{0}}= \mathcal{A} (\ket{\psi_{1\kv}}, \ldots, \ket{\psi_{m\kv}})$, with $\mathcal{A}$ an antisymmetrization over single-particle states, as a reference point for the two-parameter estimation,
we have,
\beq{}
      g^{\text{MB}}_{ij} (0) = \sum_{\kv} g_{ij}(\kv) = \frac{L_x L_y}{(2\pi)^2}\int_\text{BZ} \dd^2 \kv~g_{ij},
\eeq
with $L_x,L_y$ the linear system dimensions, and $A = L_x L_y$ the area of the material. Hence, on combining with the spin-Chern bounds of Sec.~II, we obtain:
\beq{}
   \frac{1}{4} F_{ii} [\rho]  = \text{Tr}~g^{\text{MB}}_{ii} (0)  \geq \frac{A (|C_\uparrow| +  |C_\downarrow|) }{2 \pi},
\eeq
which demonstrates the scaling of the Fisher information, and hence of the precision of the two-parameter measurements, with the size of spin-topological free fermion system. 

We now demonstrate the scaling of quantum metric with the spin gap. The eigenvalue equation for the projected spin-operator is:
\beq{}
    PS_zP \ket{u^\sigma_{n\kv}} = s^\sigma_z \ket{u^\sigma_{n\kv}}.
\eeq
On differentiating the equation, and defining the local spin gap $\Delta S_{mn} \equiv s^\uparrow_m - s^\downarrow_n$, we obtain,
\beq{}
    \bra{u^{\sigma'}_{m\kv}} \ket{\partial_{k_i} u^\sigma_{n\kv}} = \frac{\bra{u^{\sigma'}_{m\kv}} \partial_{k_i} (PS_zP) \ket{u^\sigma_{n\kv}}}{s^\sigma_n - s^{\sigma'}_m} = \frac{\bra{u^{\sigma'}_{m\kv}} \partial_{k_i} (PS_zP) \ket{u^\sigma_{n\kv}}}{\Delta S_{mn}},
\eeq
and hence for the spin-resolved quantum metric $ g^{\sigma}_{ij}$,
\beq{}
    g^{\sigma}_{ii} = \sum^{N_\text{occ}}_{n,m} \frac{\Big| \bra{u^{\sigma' \neq \sigma}_{m \kv}} \partial_{k_i} (PS_zP) \ket{u^\sigma_{n\kv}} \Big|^2}{(\Delta S_{mn})^2} + (\text{unocc}),
\eeq
where the $'\text{unocc}'$ terms correspond to the matrix elements of spin bands to unoccupied bands, which are bounded by the energy gap. We observe that as the spin gap $\Delta S_p = \text{min}~\Delta S_{mn} \rightarrow 0$ becomes vanishingly small, the spin-resolved metric scales up accordingly, as long as the matrix elements in the numerator do not decay faster than a power law. We find this scenario to occur in the context of bismuth and four-band Hamiltonians in Sec.~VIII. Correspondingly, we observe that the metric elements $g_{ii}$ scale up as the spin-gap $\Delta S_p$ decreases, which we show in the main text, and in the following section. The higher the metric, the better the sensing capability, as underpinned by the QCRBs highlighted in this Section.
 
\section{Spin-resolved quantum geometry in models realizing spin topology}\label{app::H}
\begin{figure}
    \centering
    \def\svgwidth{0.55\linewidth}
    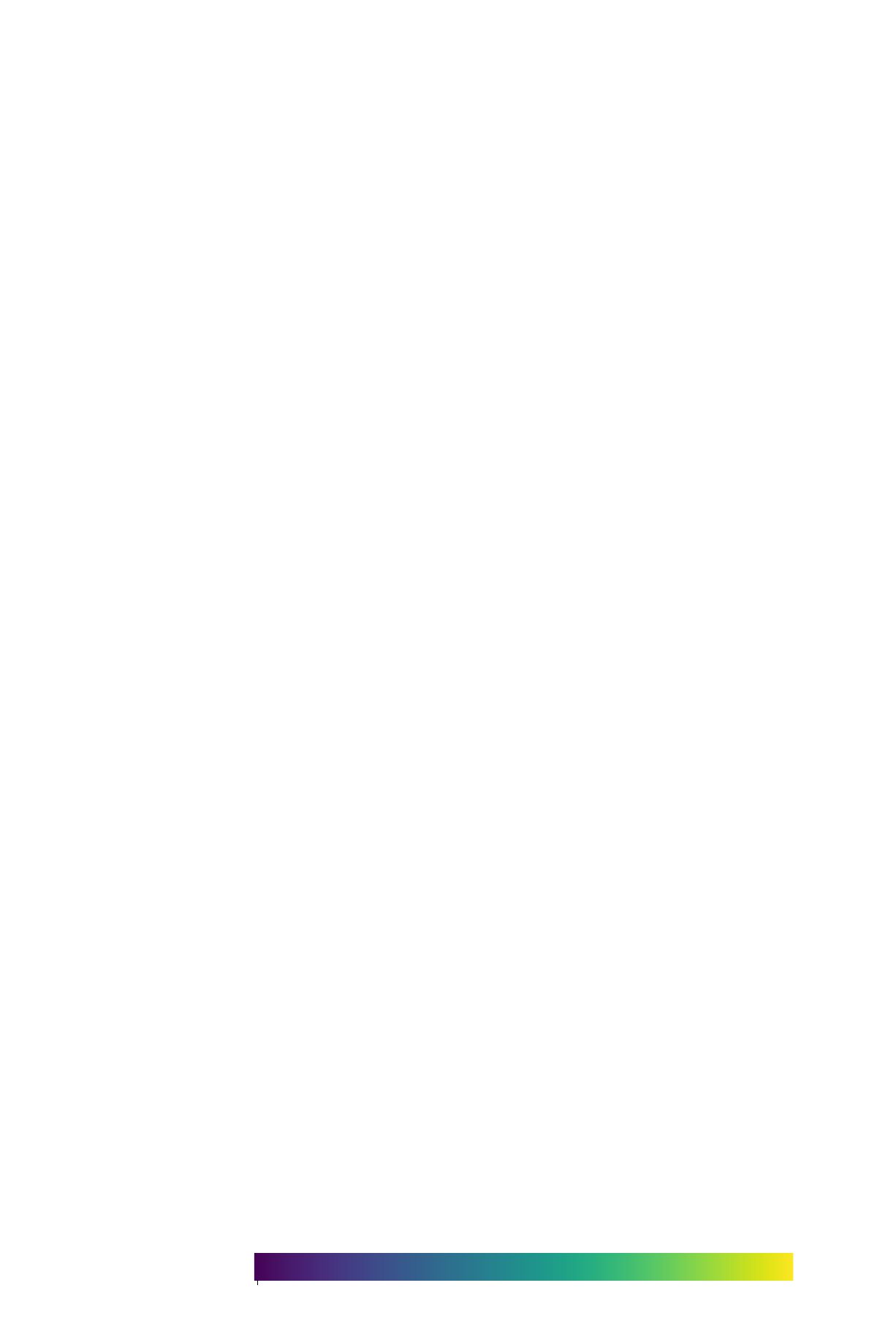
    \caption{Quantum metric for the TRS-preserving QWZ model with $C^\sigma_s = \pm2$~\,\cite{Kruthof2021} in Eq.\,\eqref{eq:minimal_model} with $\lambda = 0.5$. \textbf{(a)} Band structure, \textbf{(b)} Wilson loop, \textbf{(c)} projected spin-$z$ band structure, \textbf{(d)} projected spin-$z$ Wilson loop, \textbf{(e)} total and spin decomposed quantum metric, \textbf{(f)} total and spin decomposed Berry curvature. Integrating over the BZ in \textbf{(e)} gives $(2\pi)^{-1}\int_{\mathrm{BZ}}\ \mathrm{Tr}[g^{\uparrow}] = (2\pi)^{-1}\int_{\mathrm{BZ}}\ \mathrm{Tr}[g^{\downarrow}] = 3.20$ and $(2\pi)^{-1}\int_{\mathrm{BZ}}\ \mathrm{Tr}[g] = 6.40$. Integrating the Berry curvatures gives $(2\pi)^{-1}\int_{\mathrm{BZ}} \Omega_{xy}^{\uparrow} = -(2\pi)^{-1}\int_{\mathrm{BZ}} {\Omega_{xy}^{\downarrow} = 2.00}$ and $\int_{\mathrm{BZ}} \Omega_{zy} = 0$. }
    \label{fig:QWZ_model_topology}
\end{figure}

\begin{figure}
    \centering
    \def\svgwidth{0.55\linewidth}
    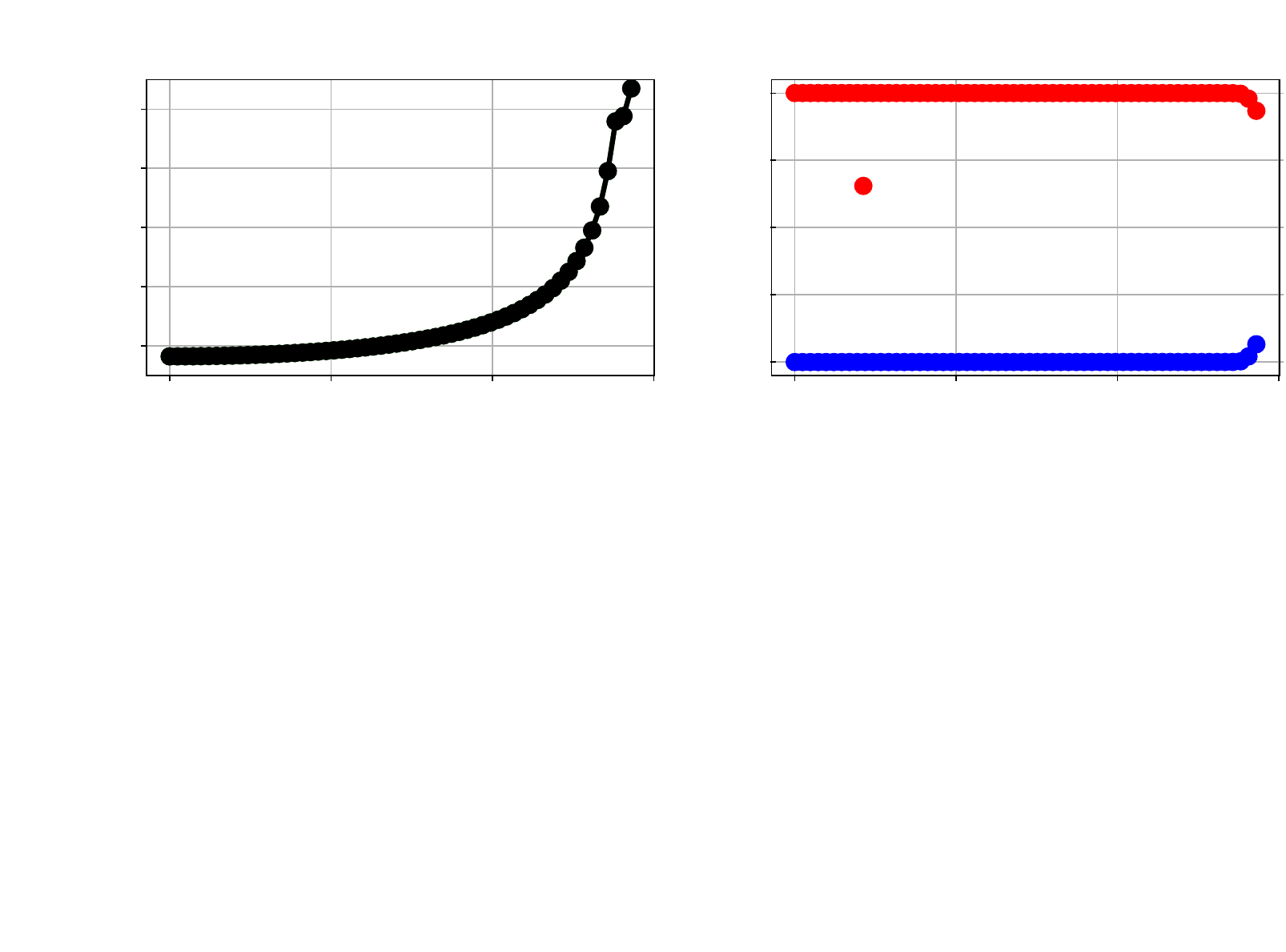
    \caption{Scaling of the TRS-symmetric QWZ model in Eq.\,\eqref{eq:minimal_model}. The spin topology \textbf{(b)} is well-defined until the energy gap closes at $\lambda \approx 0.72$ \textbf{(d)}. In the vicinity of this transition, the metric rapidly increases \textbf{(a)}, though the spin gap remains large until the energy transition \textbf{(c)}.}
     \label{fig:QWZ_model_scaling}
\end{figure}
We here detail a minimal four-band model realizing many of the same features as the ultrathin bismuth material example. This could be more easily implemented in an artificial setup such as NV-centers in diamond as in Ref.~\cite{Yu2024}. The model is based on a doubled quantum Hall model considered in Ref.~\cite{Kruthof2021}, based on a time-reversal symmetric version of the Qi-Wu-Zhang (QWZ) model \cite{QWZ_model_2006}. We consider two orbitals $s$ and $p_z$, both placed at the origin of the square unit cell with lattice parameter $a = 1$. Ordering the basis as  $\{\ket{s\uparrow},\ket{s\downarrow},\ket{p_z\uparrow},\ket{p_z\downarrow}\}$ the model is given by:
\begin{small}
\begin{equation}\label{eq:minimal_model}
H(\kv;\lambda) = \begin{pmatrix}
    m^2-(\sin k_x+\sin k_y)^2 & \lambda[\sin k_x-i\sin(k_x+k_y)] & \cos k_x-i\cos k_y & 0\\
    \lambda[\sin k_x+i\sin(k_x+k_y)] & m^2-(\sin k_x+\sin k_y)^2 & 0 & \cos k_x+i\cos k_y\\
    \cos k_x +i\cos k_y & 0& -m^2+(\sin k_x+\sin k_y)^2 &  \lambda[-\sin k_x+i\sin(k_x+k_y)]\\
    0 & \cos k_x-i\cos k_y & \lambda[-\sin k_x-i\sin(k_x+k_y)] & -m^2+(\sin k_x+\sin k_y)^2
\end{pmatrix},
\end{equation}
\end{small}
where we fix $m = 1$ throughout. Here $\lambda$ controls the SOC strength, and can be viewed as the metrological parameter. We show the metric and spin topology for this model in Fig.\,\ref{fig:QWZ_model_topology}, and scaling with $\lambda$ in Fig.\,\ref{fig:QWZ_model_scaling}. The associated optical response is shown in Fig.\,\ref{fig:Optical_response}(b). We note that in this system $g^{\uparrow}+g^{\downarrow} = g$ is satisfied for all $\lambda$ and $\kv$.

\section{Spin-resolved quantum geometry and spin topology in the presence of disorder}\label{app::I}

Below, we detail how the spin-resolved quantum geometry and its interplay with the spin topology central to this work can be characterized in the presence of disorder.  In the following, we first consider systems under open boundary conditions (OBC). We begin by defining the real-space spin-resolved metric markers and retrieve their correspondences to the localization markers capturing the quantum metric. The spin-metric marker proposed in this work reads:
\beq{}
g^\sigma_{ij}(\textbf{r}) = \frac{1}{2 A_\text{cell}} \text{Tr}_\text{cell} [P_\sigma \{P_\sigma \hat{x}_i P_\sigma, P_\sigma \hat{x}_j P_\sigma\}] = \frac{1}{2 A_\text{cell}} \sum_\alpha \bra{r_\alpha} P_\sigma \{P_\sigma \hat{x}_i P_\sigma, P_\sigma \hat{x}_j P_\sigma \} \ket{r_\alpha}.
\eeq
where $\{ \ldots, \ldots \}$ denotes an anticommutator of spin-projected operators $P_\sigma \hat{x}_i P_\sigma$, with position operator components $\hat{x}_i$ running over $i = 1,2$, corresponding to $\hat{x}$ and $\hat{y}$, respectively. $A_\text{cell}$ is an area of a unit cell. The construction of the spin-geometric marker is analogous to the construction of the localization (metric) marker introduced by Marrazzo and Resta in Ref.~\cite{Marrazzo2019}.
\beq{}
\mathcal{F}_{ij}(\textbf{r}) = - \text{Tr}_\text{cell} [P [\hat{x}_i, P], [\hat{x}_j, P]] = - \sum_\alpha \bra{r_\alpha} P [\hat{x}_i, P], [\hat{x}_j, P] \ket{r_\alpha}.
\eeq
We adapt a real-space marker for the trace of metric as:
\beq{}
    \text{Tr}~g(\textbf{r}) \equiv g_{xx}(\textbf{r}) + g_{yy}(\textbf{r}) = \frac{1}{4\pi A_\text{cell}} [ \mathcal{F}_{xx}(\textbf{r}) + \mathcal{F}_{yy}(\textbf{r}) ],
\eeq
where analogously to the spin-resolved metric markers, we include an expression equivalent to the localization marker~\cite{Chen2023, Marsal2024}:
\beq{}
g_{ij}(\textbf{r}) = \frac{1}{2 A_\text{cell}} \text{Tr}_\text{cell} [P \{P \hat{x}_i P, P \hat{x}_j P \}] = \frac{1}{2 A_\text{cell}} \sum_\alpha \bra{r_\alpha} P \{P \hat{x}_i P, P \hat{x}_j P \} \ket{r_\alpha}.
\eeq
Similarly, for the real-space spin-metric marker, we consistently define:
\beq{}
\text{Tr}~g^\sigma(\textbf{r}) \equiv g^\sigma_{xx}(\textbf{r}) + g^\sigma_{yy}(\textbf{r}). 
\eeq
The spin-Chern marker, which we compare to the spin-geometric and localization (metric) markers; and in particular to their traces, reads~\cite{Bau2024}:
\beq{}
C^{\sigma}(\textbf{r})= \frac{4\pi}{A_\text{cell}} \mathfrak{Im}~\text{Tr}_\text{cell} [P_\sigma [\hat{x}, P_\sigma], [\hat{y}, P_\sigma]] = \frac{4\pi}{A_\text{cell}} \mathfrak{Im}~ \sum_\alpha \bra{r_\alpha} P_\sigma [\hat{x}_i, P_\sigma], [\hat{x}_j, P_\sigma] \ket{r_\alpha}.
\eeq
We find that both the spin-metric and localization (metric) markers are lower-bounded by the spin-Chern marker, demonstrating that the bounds derived in this work hold locally in individual parts of disordered spin-topological systems. The validity of the local bound can be derived starting from the momentum-space inequality, as follows. We rewrite the momentum space integral as $\text{Tr}_\kv [ \ldots ] = \int_\text{BZ} \frac{\dd^2 \kv}{(2\pi)^2} (\ldots)$. Combining the integration with the local inequality Eq.~\eqref{eq:CentralBound}, we have:
\beq{}
      N_\text{occ} \sum_{i = x,y} \frac{1}{2} \text{Tr}_\kv (P \{ \partial_{k_i} P, \partial_{k_i} P \}) \equiv \text{Tr}_\kv (g_{xx} + g_{yy}) \geq \text{Tr}_\kv \sum_{\sigma = \uparrow, \downarrow} |\Om^\sigma_{xy}| = \sum_{\sigma = \uparrow, \downarrow} \text{Tr}_\kv [|P_\sigma [\partial_{k_x} P_\sigma, \partial_{k_y} P_\sigma] | ],
\eeq
where we additionally inserted definitions of the quantum metric and spin-Berry curvature in terms of projectors. Under a substitution $\partial_{k_i} \hat{P}_{(\sigma)} = i [\hat{x}_i, \hat{P}_{(\sigma)}]$, and using basis independence and completeness relations of the trace $\text{Tr}_\kv[\ldots]=\text{Tr}_\vec{r}[\ldots] = \int \dd \vec{r}~ \bra{r} \ldots \ket{r}$, we obtain:
\beq{}
      N_\text{occ} \sum_{i = x,y} \frac{1}{2} \text{Tr}_\rv (P \{ [\hat{x}_i, P], [\hat{x}_i, P] \}) \geq \sum_{\sigma = \uparrow, \downarrow} \text{Tr}_\rv [|P_\sigma [[\hat{x}, P_\sigma], [\hat{y}, P_\sigma]] |],
\eeq
which in terms of real-space markers reads:
\beq{}
    \frac{1}{2\pi} \sum_{i \in \text{cells}} \text{Tr}~g(\textbf{r}_i) \geq \sum_{i \in \text{cells}} \frac{ |C^\uparrow(\textbf{r}_i)| + |C^\downarrow(\textbf{r}_i)| }{N_\text{occ}}.
\eeq
By repeating the analogous steps, starting from inequality Eq.~\eqref{eq:SpinBound},
we have:
\beq{eq:realspacebound}
    \sum_{i \in \text{cells}} \text{Tr}~g^\sigma(\textbf{r}_i) \geq \sum_{i \in \text{cells}} 2\pi |C^\sigma(\textbf{r}_i)|. 
\eeq
As central to this work, the above real-space results show that the spin-geometric bounds also hold in the presence of disorder, as long as the spin-Chern numbers are preserved. It should be noted that the sum over all the cells in the disordered system needs to be performed, as locally:
\beq{}
    \text{Tr}~g^\sigma(\textbf{r}) = \frac{1}{2 A_\text{cell}} \sum_\alpha \bra{r_\alpha} P \{P \hat{x}_i P, P \hat{x}_j P \} \ket{r_\alpha} = \frac{1}{2 A_\text{cell}} \sum_\alpha \sum_{\vec{k}, \vec{k'}} e^{i(\vec{k} - \vec{k}') \cdot \vec{r} } \bra{\vec{k}'_\alpha} P \{P \hat{x}_i P, P \hat{x}_j P \} \ket{\vec{k}_\alpha},
\eeq
with Bloch orbitals $\ket{\vec{k}_\alpha} =  \sum_\vec{r} e^{i\vec{k} \cdot \vec{r} } \ket{r_\alpha}$, which reduces to the momentum-space bound encoding the positive-semidefiniteness of quantum-geometric tensors in the momentum parameter space only when the integration/summation $\sum_\rv e^{i(\vec{k} - \vec{k}') \cdot \vec{r}} = \delta(\vec{k} - \vec{k}')$ is performed. The real-space summation, as present in Eq.~\eqref{eq:realspacebound}, pins the matrix elements of geometric tensors to diagonal elements $\bra{\vec{k}_\alpha} P \{P \hat{x}_i P, P \hat{x}_j P \} \ket{\vec{k}_\alpha}$, which combined with the Bloch-orbital flavours $c_{n\alpha}$ directly to the eigenstates $\ket{u_{n\kv}}= \sum_\alpha c_{n\alpha} \ket{\vec{k}_\alpha}$. It should be stressed that only the corresponding elements diagonal in the eigenstate basis over the parameter space enjoy the geometric bounds, which are necessarily local in the momentum (parameter) space.

In addition to defining the spin-resolved geometric markers under the OBC, we consider the spin-resolved geometric markers for disordered systems under periodic boundary conditions (PBC). The PBC spin-resolved metric marker reads,
\beq{}
[g^\sigma_{ij}(\textbf{r})]_\text{PBC} = \frac{1}{2 A_\text{cell}} \text{Tr}_\text{cell} [P^\sigma_\Gamma \{P^\sigma_{\vec{b}_i}, P^\sigma_{\vec{b}_j}\}] = \frac{1}{2 A_\text{cell}} \sum_\alpha \bra{r_\alpha} P^\sigma_\Gamma \{P^\sigma_{\vec{b}_i}, P^\sigma_{\vec{b}_j} \} \ket{r_\alpha}.
\eeq
with the PBC metric marker:
\beq{}
[g_{ij}(\textbf{r})]_\text{PBC} = \frac{1}{2 A_\text{cell}} \text{Tr}_\text{cell} [P_\Gamma \{P_{\vec{b}_i}, P_{\vec{b}_j} \} ] = \frac{1}{2 A_\text{cell}} \sum_\alpha \bra{r_\alpha} P_\Gamma \{P_{\vec{b}_i}, P_{\vec{b}_j} \} \ket{r_\alpha},
\eeq
and the PBC spin-Chern marker, as originally defined in Ref.~\cite{Bau2024}:
\beq{}
[C_{\sigma}(\textbf{r})]_\text{PBC}= \frac{4\pi}{A_\text{cell}} \mathfrak{Im}~\text{Tr}_\text{cell} [P^\sigma_\Gamma [P^\sigma_{\vec{b}_1}, P^\sigma_{\vec{b}_2}] ] = \frac{4\pi}{A_\text{cell}} \mathfrak{Im}~ \sum_\alpha \bra{r_\alpha} P^\sigma_\Gamma [P^\sigma_{\vec{b}_1}, P^\sigma_{\vec{b}_2}] \ket{r_\alpha}.
\eeq
Here, the PBC projectors are defined within an approximation of Ref.~\cite{Bau2024}:
\beq{}
 P^{(\sigma)}_{\vec{b}_i} \approx P^{(\sigma)}_\Gamma + \frac{2\pi i}{L} [\hat{x}_i, P^{(\sigma)}_\Gamma] - \frac{2\pi^2}{L^2} [\hat{x}_i, [\hat{x}_i, P^{(\sigma)}_\Gamma]],
\eeq
where $L$ is the linear system size, and the $P^\sigma_\Gamma$ projector reads~\cite{Bau2024}:
\beq{}
    P^\sigma_\Gamma \equiv \sum^{N_\text{occ}/2}_n \ket{u^\sigma_{n{\Gamma}}} \bra{u^\sigma_{n{\Gamma}}}. 
\eeq
In the case of the single-point metric, analogously~\cite{Bau2024Chern}:
\beq{}
    P_\Gamma \equiv \sum^{N}_n \ket{u_{n{\Gamma}}} \bra{u_{n{\Gamma}}}. 
\eeq
In the above, $\ket{u_{n{\Gamma}}}$ and $\ket{u^\sigma_{n{\Gamma}}}$ denote the  $\Gamma$-point eigenstates and spin-eigenstates, respectively, i.e., all the occupied zero-momentum (spin-)eigenstates within a supercell BZ shrunk to a point in the aperiodic limit defined in the presence of disorder, as introduced in Refs.~\cite{Ceresoli2007, Favata2023}. The PBC formulation of the markers with the above projectors is consistent with the single-point formulation of topological invariants, as introduced in Ref.~\cite{Ceresoli2007} for the standard Chern numbers. 

Furthermore, to characterize spin topology and spin-resolved quantum geometry under disorder, we define the single-point (spin-)metric estimators, consistently with the single-point Chern and spin-Chern invariants \cite{Ceresoli2007,Favata2023}. We demonstrate that the single-point metric estimators satisfy the geometric bounds even in the presence of disorder, demonstrating the robustness of the spin-geometric bound up to the spin-gap closing transition. The single-point spin-resolved metric reads:
\beq{}
[g^\sigma_{ij}]_{\text{sp}} = \text{Tr}_\rv [P^\sigma_\Gamma \{P^\sigma_{\vec{b}_i}, P^\sigma_{\vec{b}_j}\}],
\eeq
with the single-point quantum metric estimator:
\beq{}
[g_{ij}]_{\text{sp}} = \text{Tr}_\rv [P_\Gamma \{P_{\vec{b}_i}, P_{\vec{b}_j}\}],
\eeq
and the single-point spin-Chern invariant~\cite{Favata2023} satisfying a lower bound on the employed single-point metrics defined above:
\beq{}
[C^{\sigma}_s]_{\text{sp}} = 4\pi~ \mathfrak{Im}~\text{Tr}_\rv [P^\sigma_\Gamma [P^\sigma_{\vec{b}_1}, P^\sigma_{\vec{b}_2}] ].
\eeq
The single-point bounds for the disordered systems  can be expressed as, 
\beq{}
\text{Tr}~[g^\sigma_{ij}]_{\text{sp}} \geq 2\pi [C^{\sigma}_s]_{\text{sp}},
\eeq
for the single-point spin-resolved quantum metric, and for the single-point quantum metric,
\beq{}
N_\text{occ}\text{Tr}~[g_{ij}]_{\text{sp}} \geq 2\pi \sum_{\sigma = \uparrow, \downarrow} [C^{\sigma}_s]_{\text{sp}}.
\eeq
The corresponding proofs follow by the same logical argument, analogously to the proof of inequality Eq.~\eqref{eq:realspacebound}, on connecting the PBC projectors to the momentum-space projectors within the approximations of Refs.~\cite{Ceresoli2007, Favata2023, Bau2024}.

In the following Section, we detail the computational evaluation of the here-defined spin-resolved quantum geometric markers in real space, as applicable under disorder naturally present in real spin-topological materials. 

\section{Further computational details on disorder calculations}\label{app::J}

In the following, we provide further details on disorder calculations in spin-topological systems considered in this work. We consider random potential (Anderson) disorders with strengths $W$, which we include in the real-space Hamiltonians of form,
\beq{}
    H_W = \sum_{\alpha, \vec{r}} \delta \mu_{\alpha}(\vec{r}) \hat{c}^\dagger_{\alpha,\vec{r}} \hat{c}_{\alpha, \vec{r}}, 
\eeq
where $\hat{c}^\dagger_{\vec{r}, \alpha}$ and $\hat{c}^\dagger_{\vec{r}, \alpha}$ are respectively the creation and annihilation operators for an electron in orbital $\alpha$ at the unit cell at the position vector $\vec{r}$. The random potential (Anderson) disorder satisfies a uniform probability distribution for orbital-dependent local onsite potentials $\delta \mu_{\alpha}(\vec{r}) \in [-W, W]$. We introduce the random potential within the tight-binding framework of the \texttt{StraWBerryPy} code \cite{Bau2024Chern, Bau2024}, with the real-space tight-binding Hamiltonian parameters implemented explicitly for the four-band (TRS-QWZ) Hamiltonian, Eq.~\eqref{eq:minimal_model}, and the tight-binding Hamiltonian parameters for Bi automatically read out from the \texttt{Wannnier90} output. The combined disordered Hamiltonian reads
\beq{}
    H = \sum_{\alpha, \vec{r}} \varepsilon_\alpha \hat{c}^\dagger_{\alpha,\vec{r}} \hat{c}_{\alpha, \vec{r}} + \sum_{\alpha, \beta, \vec{r}, \vec{r}'} t_{\alpha \beta}(\vec{r} - \vec{r}') \hat{c}^\dagger_{\alpha,\vec{r}} \hat{c}_{\beta, \vec{r}'} + H_W, 
\eeq
with the original orbital onsite energies $\varepsilon_{\alpha}$ and the real-space hoppings $t_{\alpha \beta}(\vec{r} - \vec{r}')$ of the clean, i.e., non-disordered, Hamiltonians.

\begin{figure}[t!]
    \centering
    \def\svgwidth{0.75\linewidth}
    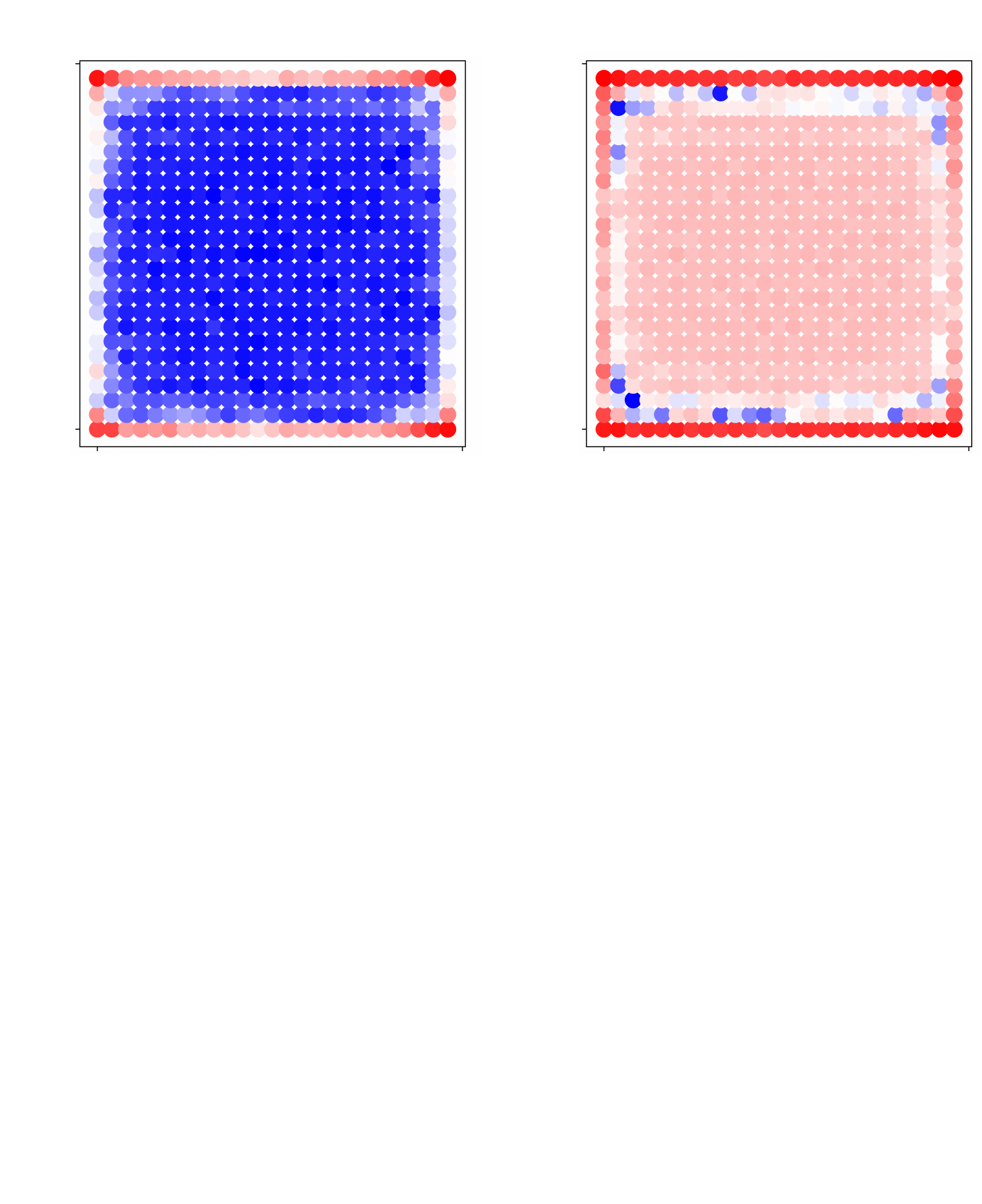
    \caption{Robustness of spin-topologically bounded quantum geometry in the presence of disorder in the four-band model in Eq.~\eqref{eq:minimal_model} with $C_\sigma = \pm2$. \textbf{(a)} Spin-resolved metric markers $\frac{1}{2\pi} \text{Tr}[g^{\sigma}(\vec{r})]$ and \textbf{(b)} spin-Chern markers $C^\sigma(\vec{r})$ in a disordered system of $625$ atoms, with random potential disorder of strength ${W = 1.6~\text{eV}}$. Averaging over 81 atoms in the center yields: ${\frac{1}{2\pi} \overline{\text{Tr}~g^{\uparrow}(\vec{r})}=3.25}$ and ${\overline{C^{\uparrow}(\boldsymbol{r})} = 2.00}$. \textbf{(c)} Scaling of single-point metrics $g^{(\sigma)}_{\text{sp}} \equiv \frac{1}{2\pi} \text{Tr}[{g}^{(\sigma)}_{ij}]_\text{sp}$, \textbf{(d)} single-point spin-Chern numbers $C^\sigma_\text{sp} \equiv [C^\sigma_s]_\text{sp}$, \textbf{(e)}~spin gap $\Delta S_p$, and \textbf{(f)} energy gap $\Delta $E against disorder strength $W$ in a system of 625 atoms, averaged over 20 disorder realizations. The real space markers are averaged over the central 225 atoms.} 
    \label{fig:FigDisorder4band}
\end{figure}

In addition to the disorder calculations for the ultrathin Bi systems with $C_s = 2$, we here include the analogous calculations in a minimal four-band model with $C_s = 2$, see Fig.~\ref{fig:FigDisorder4band} for reference. We find that the (spin-)geometric bounds reformulated in the real-space formalism in the previous Section (Sec.~IX) hold in the four-band model, analogously to the real material realization of ultrathin Bi demonstrated in the main text.

In the case of the four-band model, the real-space hoppings obtained on Fourier-transforming the momentum-space matrix elements of the Bloch Hamiltonian~Eq.~\eqref{eq:minimal_model} expressed in the orbital basis, $\alpha \in \{\ket{s\uparrow},\ket{s\downarrow},\ket{p_z\uparrow},\ket{p_z\downarrow}\}$,  take the following parametric values:
\beq{}
    t_{s\uparrow p\uparrow}(1,0) = t_{s\uparrow p\uparrow}(1,0) = t_{s\downarrow p\downarrow}(1,0) = t_{s\downarrow p\downarrow}(-1,0) = \frac{1}{2}, 
\eeq
\beq{}
    t_{s\downarrow p\downarrow}(0,1) = t_{s\downarrow p\downarrow}(0,-1) = -t_{s\uparrow p\uparrow}(0,1) = -t_{s\uparrow p\uparrow}(0,-1) = \frac{i}{2}, 
\eeq
\beq{}
    t_{s\uparrow s\downarrow}(1,0) =  t_{p\uparrow p\downarrow}(1,0) = -t_{s\uparrow s\downarrow}(-1,0) =  -t_{p\uparrow p\downarrow}(-1,0) = \frac{i \lambda}{2},
\eeq
\beq{}
    t_{s\uparrow s\downarrow}(-1,-1) =  t_{p\uparrow p\downarrow}(1,1) =  -t_{s\uparrow s\downarrow}(1,1) = -t_{p\uparrow p\downarrow}(-1,-1) = \frac{\lambda}{2},
\eeq
\beq{}
    t_{s\uparrow s\uparrow}(1,-1) = t_{s\uparrow s\uparrow}(-1,1) = t_{s\downarrow s\downarrow}(1,-1) = t_{s\downarrow s\downarrow}(-1,1)  = t_{p\uparrow p\uparrow}(1,1) = t_{p\uparrow p\uparrow}(-1,-1) = t_{p\downarrow p\downarrow}(1,1) = t_{p\downarrow p\downarrow}(-1,-1) = i, 
\eeq
\beq{}
    t_{s\uparrow s\uparrow}(1,1) = t_{s\uparrow s\uparrow}(-1,-1) = t_{s\downarrow s\downarrow}(1,1) = t_{s\downarrow s\downarrow}(-1,-1) = t_{p\uparrow p\uparrow}(1,-1) = t_{p\uparrow p\uparrow}(-1,1) = t_{p\downarrow p\downarrow}(1,-1) = t_{p\downarrow p\downarrow}(-1,1) = -i, 
\eeq
\beq{}
    t_{p\uparrow p\uparrow}(0,\pm 2) = t_{p\uparrow p\uparrow}(\pm 2, 0) = t_{p\downarrow p\downarrow}(0,\pm 2) = t_{p\downarrow p\downarrow}(\pm 2, 0) =  \frac{1}{2}, 
\eeq
\beq{}
 t_{s\uparrow s\uparrow}(0,\pm 2) = t_{s\uparrow s\uparrow}(\pm 2, 0) = t_{s\downarrow s\downarrow}(0,\pm 2) = t_{s\downarrow s\downarrow}(\pm 2, 0) = -\frac{1}{2},
\eeq
with the onsite energies amounting to:
\beq{}
    \varepsilon_{s\uparrow} = \varepsilon_{s\downarrow} = -\varepsilon_{p\uparrow} = -\varepsilon_{p\uparrow} = -m^2 + 1.
\eeq
We adapt spin-Chern marker and single-points spin-Chern number formulation of Refs.~\cite{Favata2023, Bau2024} in the \texttt{StraWBerryPy} implementation. By replacing the imaginary parts of position basis matrix elements with real parts, and changing the projectors for spin-projectors, we extend the formulation of the localizer~\cite{Marrazzo2019} to the spin-resolved metric markers defined in the previous Section. Analogously, we construct the single-point spin-metric estimator, by replacing the projectors of the single-point spin-Chern numbers implemented by Ref.~\cite{Favata2023}. 

Consistently with the convergence within a finite-size scaling study of the single-point spin-Chern invariants and spin-Chern markers of Refs.~\cite{Favata2023, Bau2024}, we adapt system sizes with linear dimensions $L = 15-21$ unit cells of $4$ Bi atoms ($24$ orbitals) to correspondingly model realistic Bi nanoflakes ($\sim2000$ atoms, $>10,000$ orbitals), while minimizing the computational cost. We find that the results for different system sizes are in agreement within the numerical errors of the previously introduced methods~\cite{Favata2023, Bau2024}.
\begin{figure}[t!]
    \centering
    \def\svgwidth{0.75\linewidth}
    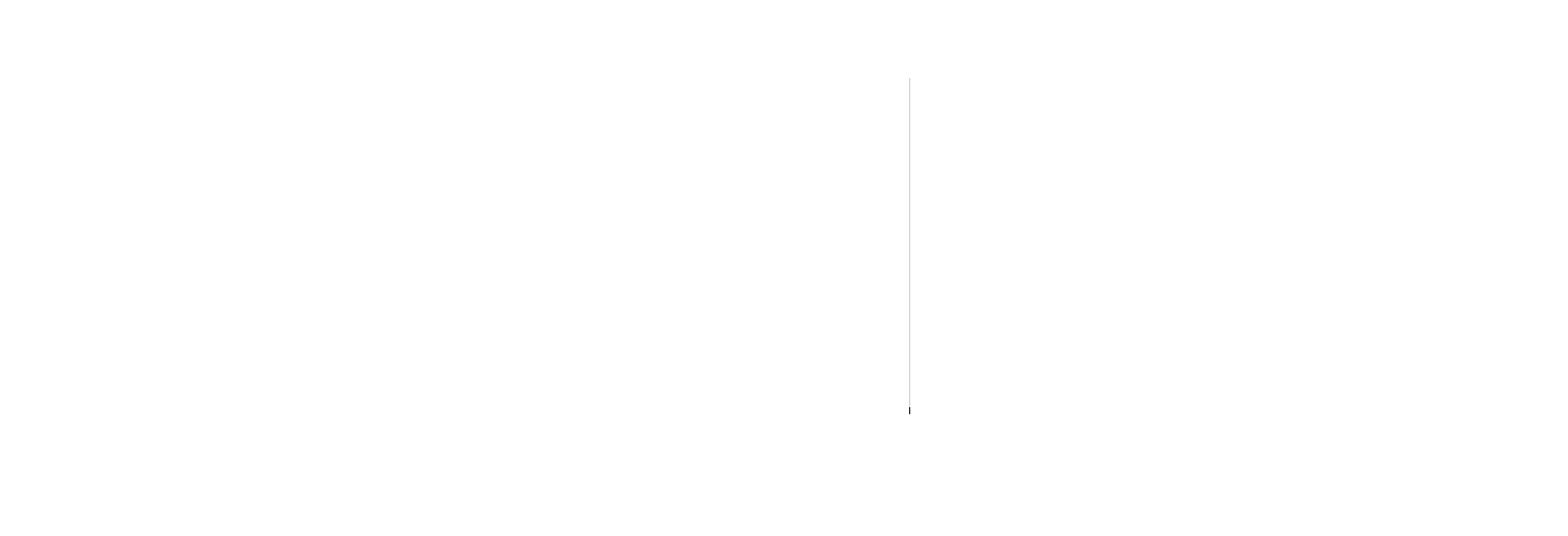
    \caption{Dissolution of spin-topologically bounded quantum geometry in the presence of high disorder in the Anderson localized regime. In the high disorder limit, the (single-point) spin-Chern numbers vanish, and the spin topology is trivial, which allows for the vanishing of single-point (spin-)metric and metric markers. We observe that the single-point (spin-)metric vanishes in the high disorder limit for both the {\textbf{(a)}} ultrathin Bi and {\textbf{(b)} minimal four-band model realizations of the $\mathbb{Z}_2$-trivial spin-Chern numbers $C_s = 2$. For the four-band model, we additionally find that the averaged (spin-)metric markers vanish analogously, while the initial divergences of the (spin-)metric are suppressed due to the finite size of the system.}} 
    \label{fig:FigDisorderHigh}
\end{figure}

Finally, beyond the limit of subcritical disorders introduced in the clean insulating material, it is furthermore interesting to observe the behaviour of the (spin-resolved) quantum geometry in the highly-disordered regimes admitting Anderson localization of the metallic phases realizable with lower disorder strengths. The corresponding localization transitions follow subsequently to the spin-gap and energy gap-closing demonstrated in the main text, and after the delocalization transition to a metallic state and associated trivialization of the spin-Chern numbers occurs. Below, we correspondingly demonstrate real-space marker and single-point estimator results for both ultrathin Bi and the four-band model in the limit of the strong disorder exceeding the size of the bandwidth. We find that the (spin-resolved) quantum metric approaches zero, consistently with being lower-bounded by trivialized $C_s = 0$ in the Anderson-localized regime at high disorder. We show the corresponding dissolution of the real-space (spin-resolved) quantum metric in the Anderson insulator state expected in the high disorder regime, see Fig.~\ref{fig:FigDisorderHigh} for reference.

\end{widetext}

\clearpage

\section*{Supplementary References}
\bibliographystyle{apsrev4-1}
\bibliography{references.bib}